%% file: ANA-HDBS-2021-17-PAPER.tex
\pdfoutput=1
\documentclass[cernpreprint, atlasdraft=false, texlive=2023, UKenglish, texmf, orcidlogo]{atlasdoc}
\usepackage{atlaspackage}
\usepackage{atlasbiblatex}

\usepackage{atlasphysics}
\graphicspath{{logos/}{figures/}}

\usepackage{ANA-HDBS-2021-17-PAPER-defs}

\newcommand{\AtlasCoordFootnote}{%
ATLAS uses a right-handed coordinate system with its origin at the nominal interaction point (IP)
in the centre of the detector and the \(z\)-axis along the beam pipe.
The \(x\)-axis points from the IP to the centre of the LHC ring,
and the \(y\)-axis points upwards.
Polar coordinates \((r,\phi)\) are used in the transverse plane,
\(\phi\) being the azimuthal angle around the \(z\)-axis.
The pseudorapidity is defined in terms of the polar angle \(\theta\) as \(\eta = -\ln \tan(\theta/2)\) and is equal to the rapidity
$ y = \frac{1}{2} \ln \left( \frac{E + p_z c}{E - p_z c} \right) $ in the relativistic limit.
Angular distance is measured in units of \(\Delta R \equiv \sqrt{(\Delta y)^{2} + (\Delta\phi)^{2}}\).}

\AtlasTitle{Search for a resonance decaying into a scalar particle and a Higgs boson in the final state with two bottom quarks and two photons in proton--proton collisions at $\sqrt{s}$=13~TeV with the ATLAS detector}

\AtlasAbstract{%
A search for the resonant production of a heavy scalar $\ensuremath{X}$ decaying into a Higgs boson and a new lighter scalar
${\ensuremath{S}}$, through the process $\ensuremath{X} \to{\ensuremath{S({\to}b\bar b})\xspace}{\ensuremath{H({\to}\gamma\gamma})\xspace}$,
where the two photons are consistent with the Higgs boson decay, is
performed. The search is conducted using an integrated luminosity of 140~fb$^{-1}$ of proton-proton collision data at
a centre-of-mass energy of 13~TeV recorded with the ATLAS detector at the Large Hadron Collider. The search is
performed over the mass range 170~$\leq$~${\ensuremath{m_{X}}\xspace}$~$\leq$~1000~GeV and 15~$\leq$~${\ensuremath{m_{S}}\xspace}$~$\leq$~500~GeV.
Parameterised neural networks are used to enhance the signal purity and to achieve continuous sensitivity in a
domain of the (${\ensuremath{m_{X}}\xspace}$, ${\ensuremath{m_{S}}\xspace}$) plane. No significant excess above the
expected background is found and 95\% CL upper
limits are set on the cross section times branching ratio, ranging from 39~fb to 0.09~fb. The largest deviation
from the background-only expectation occurs for  (${\ensuremath{m_{X}}\xspace}$, ${\ensuremath{m_{S}}\xspace}$) = (575, 200)~GeV with a local (global) significance
of 3.5 (2.0) standard deviations.}

\AtlasRefCode{HDBS-2021-17}

\PreprintIdNumber{CERN-EP-2024-072}

\AtlasJournal{JHEP}
\AtlasJournalRef{JHEP 11 (2024) 047}
\AtlasDOI{10.1007/JHEP11(2024)047}

\hypersetup{pdftitle={ATLAS document},pdfauthor={The ATLAS Collaboration}}

\begin{document}

\maketitle

\section{Introduction}
%


The properties of the Higgs boson ($\hsm$) discovered in 2012~\cite{HIGG-2012-27,CMS-HIG-12-028} by the
ATLAS and CMS experiments at the Large Hadron Collider (LHC) are consistent with the Standard Model (SM)
predictions~\cite{CMS-HIG-22-001-with-err-ref,HIGG-2021-23-with-err-ref}.
However the current experimental precision does not exclude that $\hsm$ may have a small mixing with
additional scalar bosons, and may be part of an extended Higgs sector. Many Beyond the Standard Model
(BSM) theories predict such an extended Higgs sector, where one of the physical Higgs boson states could
correspond to the spin-0 boson observed with a mass of 125~\GeV, while additional scalars remain to be
discovered~\cite{Ferrari:2021hqc}.

In this paper, the complete proton-proton dataset collected by the ATLAS experiment during Run~2 of the
LHC is used to search for two additional scalar bosons $\X$ and $\S$.
Under the condition $\mX>\mS+\mHiggs$, the decay $\X \to \S \hsm$ is kinematically allowed.
This phenomenology may arise in models where the SM Higgs sector is extended with either a complex
singlet~\cite{PhysRevD.79.015018} or two real singlets~\cite{TRSM}, and in models such as the complex
two-Higgs-doublet model (2HDM)~\cite{ginzburg2002twohiggsdoublet}, the 2HDM extended by a real scalar singlet~\cite{PhysRevD79023521,Muhlleitner2017}
or the Next-to-Minimal Supersymmetric Standard Model~\cite{ELLWANGER20101,ManiatisNMSSM}.

The sensitivity of the LHC to the decay $\X \to \S \hsm$ has been explored in several benchmark
scenarios~\cite{PhysRevD.99.055048,NMSSM_2HDMS_atlhc,TRSM}. The decay of the scalar $\S$ is model-
and mass-dependent. The CMS Collaboration has performed searches for $\X\to \Sbb \Hbb$, $\X\to \Sbb \Htautau$
and $\X\to \Sbb \Hyy$ using Run 2 data~\cite{CMS-B2G-21-003,JHEP11.2021.057,cms_shbbyy}. In the
$\X\to \Sbb \Hyy$ final state, CMS observed a deviation from the background-only hypothesis with a local
(global) significance of 3.8 (2.8) standard deviations at \mX=~650~\GeV\ and \mS=~90~\GeV. ATLAS
published results on the search for $\X\to \SVV \Htautau$, where $V$ denotes a $W$ or $Z$ boson~\cite{ATLAS-SHlltautau}.

\begin{figure}[tbp]
\centering
\subfloat{\includegraphics[width=0.45\textwidth]{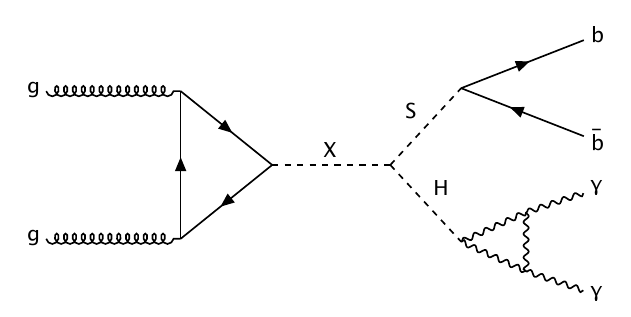}}
\caption{Example of a Feynman diagram showing gluon--gluon fusion production of a scalar $X$ decaying into a scalar $S$ and a Standard Model Higgs boson, which in turn decay into a pair of $b$-quarks and a pair of photons.}
\label{fig:feynman}
\end{figure}

This paper is focused on the search for $\X\to \Sbb \Hyy$ and uses the same Run 2 dataset already exploited by ATLAS to
search for Higgs boson pair production~\cite{HDBS-2018-34}. A di-photon mass peak arises from $\hsm \to \gamma \gamma$,
while the two $b$-tagged jets arise from the \Stobb decay, thus leading to a characteristic signal with three
resonant mass peaks from $\hsm \to \gamma \gamma$, \Stobb and $\X\to b\bar{b}\gamma\gamma$. The natural widths of the
new bosons are assumed to be much smaller than the experimental resolutions. In the particular scenario where the
scalar $\S$ has couplings similar to those of the SM Higgs boson, \Stobb is the predominant decay for $\mS<$~130~\GeV.
The Feynman diagram for the main production mode of this process is illustrated in Figure~\ref{fig:feynman}.

The rate of production for the scalar $\X$, and the decay branching ratios $BR(\X \to \S \hsm)$ and
$BR(\Stobb)$, are strongly dependent on which model is realised, and on the specific values of the
parameters of the extended Higgs sector. Therefore, the results are expressed as 95\% confidence level (CL)
upper limits on $\sigma(pp\to\X)\times BR(\X\to\S\hsm)\times BR(\Stobb)\times BR(\hsm\to \gamma\gamma)$, denoted
$\sigma(\X\to\S\hsm\to b\bar{b}\gamma\gamma)$, rather than on specific models, probing the \mX\ range
between 170 and 1000~\GeV\ and the \mS\ range between 15 and 500~\GeV.

This article is structured as follows. Section~\ref{sec:detector} briefly introduces the ATLAS detector.
Data and simulated event samples are given in Section~\ref{sec:samples}. Object definitions are introduced
in Section~\ref{sec:object_definitions} while the event selection is described in
Section~\ref{sec:event_selection}. The strategy for background estimation is explained in
Section~\ref{sec:background_estimation}. Section~\ref{sec:systematics} is devoted to the description of
the systematic uncertainties. Statistical modelling, validation of the background calculations and results
are presented in Section~\ref{sec:results}. Finally the conclusions are given in Section~\ref{sec:conclusion}.


%

%
\section{ATLAS detector}
\label{sec:detector}


%
The ATLAS detector~\cite{PERF-2007-01} at the LHC covers nearly the entire solid angle around the collision point.\footnote{\AtlasCoordFootnote}
It consists of an inner tracking detector surrounded by a thin superconducting solenoid, electromagnetic and hadronic calorimeters,
and a muon spectrometer incorporating three large superconducting air-core toroidal magnets.

The inner-detector system (ID) is immersed in a \qty{2}{\tesla} axial magnetic field
and provides charged-particle tracking in the range \(|\eta| < 2.5\).
The high-granularity silicon pixel detector covers the vertex region and typically provides four measurements per track,
the first hit generally being in the insertable B-layer (IBL) installed before Run~2~\cite{ATLAS-TDR-19,PIX-2018-001}.
It is followed by the SemiConductor Tracker (SCT), which usually provides eight measurements per track.
These silicon detectors are complemented by the transition radiation tracker (TRT),
which enables radially extended track reconstruction up to \(|\eta| = 2.0\).
The TRT also provides electron identification information
based on the fraction of hits (typically 30 in total) above a higher energy-deposit threshold corresponding to transition radiation.

The calorimeter system covers the pseudorapidity range \(|\eta| < 4.9\). Within the region \(|\eta|< 3.2\),
electromagnetic calorimetry is provided by a lead/liquid-argon (LAr) sampling calorimeter with accordion geometry.
It is divided into a barrel section covering \(|\eta| < 1.475\) and two endcap sections covering \(1.375 < |\eta| < 3.2\).
For \(|\eta| < 2.5\), it is divided into three layers in depth, which are finely segmented in $\eta$ and $\phi$.
An additional thin LAr presampler layer covering \(|\eta| < 1.8\) is used to correct for energy loss in material
upstream of the calorimeters. Hadronic calorimetry is provided by the steel/scintillator-tile calorimeter,
segmented into three barrel structures within \(|\eta| < 1.7\), and two copper/LAr hadronic endcap calorimeters.
The solid angle coverage is completed with forward copper/LAr and tungsten/LAr calorimeter modules
optimised for electromagnetic and hadronic energy measurements respectively.

The muon spectrometer (MS) comprises separate trigger and
high-precision tracking chambers measuring the deflection of muons in a magnetic field generated by the superconducting air-core toroidal magnets.
The field integral of the toroids ranges between \num{2.0} and \qty{6.0}{\tesla\metre}
across most of the detector.
Three layers of precision chambers, each consisting of layers of monitored drift tubes, cover the region \(|\eta| < 2.7\),
complemented by cathode-strip chambers in the forward region, where the background is highest.
The muon trigger system covers the range \(|\eta| < 2.4\) with resistive-plate chambers in the barrel, and thin-gap chambers in the endcap regions.

The luminosity is measured mainly by the LUCID--2~\cite{LUCID2} detector that records Cherenkov light produced in the quartz windows of photomultipliers located close to the beampipe.

Events are selected by the first-level trigger system implemented in custom hardware,
followed by selections made by algorithms implemented in software in the high-level trigger~\cite{TRIG-2016-01}.
The first-level trigger accepts events from the \qty{40}{\MHz} bunch crossings at a rate below \qty{100}{\kHz},
which the high-level trigger further reduces in order to record complete events to disk at about \qty{1}{\kHz}.

A software suite~\cite{ATL-SOFT-PUB-2021-001} is used in data simulation, in the reconstruction
and analysis of real and simulated data, in detector operations, and in the trigger and data acquisition
systems of the experiment.


%

%
\section{Data and simulated event samples}
\label{sec:samples}

The data used in this search were collected by the ATLAS experiment between 2015 and 2018, from
proton-proton collisions at $\sqrt{s}$ = 13~\TeV\ at the LHC. After data quality requirements~\cite{DAPR-2018-01},
this corresponds to an integrated luminosity of \ourLumi. The uncertainty in the combined 2015--2018 integrated luminosity is 0.83\%~\cite{FinalRun2Luminosity}, obtained using the LUCID-2 detector~\cite{LUCID2} for primary luminosity measurements, complemented by measurements using the inner detector and calorimeters.

Events are recorded using diphoton triggers that require two reconstructed photon candidates with minimum
transverse energies of 35~\GeV\ and 25~\GeV\ \cite{TRIG-2018-05}. The triggers used in 2015 and 2016 require
both the photons to satisfy the {\em Loose} photon identification criterion defined in Ref.~\cite{EGAM-2018-01}, while
the {\em Medium} criterion~\cite{EGAM-2018-01} is used for 2017–2018 to cope with the increased \pp\ interaction rate.

The Monte Carlo (MC) simulated event samples used in the analysis are listed in Table~\ref{tab:mc_samples}, along with the
generator used in the simulation, the parton distribution function (PDF) set, the showering model and the set of tuned parameters (tune).
The $\X \to \S \hsm$ signal process is simulated at leading-order (LO) in QCD with \PYTHIA[8.2]~\cite{Sjostrand:2014zea}. The
Higgs boson is forced to decay into two photons, while the scalar \S\ is forced to decay into two $b$-quarks.
The \X\ and \S\ scalar decays are generated in the narrow-width approximation. A total of 161 signal mass points
are generated in the range 170~$\leq$~\mX~$\leq$~1000~\GeV\ and 15~$\leq$~\mS~$\leq$~500~\GeV.

\begin{table}[t]
\centering
\caption{Summary of the main signal and background samples, split by production mode: signal samples, continuum background samples, single Higgs boson processes and Higgs-boson pair production samples. The generator used in the simulation, the PDF set, the showering model and the set of tuned parameters are also provided.
\label{tab:mc_samples}}
\resizebox{\columnwidth}{!}{%
\begin{tabular}{ lllll }
\toprule
\textbf{Process}         & \textbf{Generator}                           & \textbf{PDF set}           & \textbf{Showering}    & \textbf{Tune} \\
\midrule
$\X \to \S \hsm$         & \PYTHIA[8.2]~\cite{Sjostrand:2014zea} & \NNPDF[2.3lo]~\cite{Ball:2012cx} & \PYTHIA[8.2]~\cite{Sjostrand:2014zea} & A14~\cite{ATL-PHYS-PUB-2014-021} \\
\midrule
$\gamma\gamma+$jets      & \SHERPA[2.2.4]~\cite{Bothmann:2019yzt} & \NNPDF[3.0nnlo]~\cite{Ball:2014uwa}  &  -- & -- \\
$t\bar{t} \gamma \gamma$ & \MGNLO~\cite{Alwall:2014hca}  &  \NNPDF[2.3lo] & \PYTHIA[8.2] & A14 \\
$Z({\to}q\bar{q}) \gamma \gamma$ & \SHERPA 2.2.11~\cite{Bothmann:2019yzt} & \NNPDF[3.0nnlo] &  -- & -- \\
\midrule
ggF $H$          & \AtlasMC{Nnlops}~\cite{Hamilton:2012np,Campbell:2012am,Hamilton:2012rf}~\cite{Bozzi:2005wk,deFlorian:2011xf} & PDF4LHC15~\cite{Butterworth:2015oua}&   \PYTHIA[8.2]  &  AZNLO~\cite{Aad:2014xaa}      \\
VBF $H$          & \POWHEGBOX[v2]~\cite{Nason:2004rx, Frixione:2007vw, Alioli:2010xd, Nason:2009ai} & PDF4LHC15       & \PYTHIA[8.2] & AZNLO  \\
$WH$             & \POWHEGBOX[v2]~\cite{Mimasu:2015nqa, Luisoni:2013kna}                            & PDF4LHC15       & \PYTHIA[8.2] & AZNLO    \\
$qq\to ZH$       & \POWHEGBOX[v2]~\cite{Mimasu:2015nqa, Luisoni:2013kna}                            & PDF4LHC15       & \PYTHIA[8.2] & AZNLO    \\
$gg\to ZH$       & \POWHEGBOX[v2]~\cite{Mimasu:2015nqa, Luisoni:2013kna}                            & PDF4LHC15       & \PYTHIA[8.2] & AZNLO   \\
$t\bar{t}H$      & \POWHEGBOX[v2]~\cite{Hartanto:2015uka}                                           & \NNPDF[3.0nlo]  & \PYTHIA[8.2] & A14  \\
$b\bar{b}H$      & \POWHEGBOX[v2]~\cite{Campbell:2012am}                                            & \NNPDF[3.0nlo]  & \PYTHIA[8.2] & A14   \\
$tHq$            & \MGNLO                                                                           & \NNPDF[3.0nlo]  & \PYTHIA[8.2] & A14    \\
$tHW$            & \MGNLO                                                                           & \NNPDF[3.0nlo]  & \PYTHIA[8.2] & A14   \\
\midrule
ggF $HH$         & \POWHEGBOX[v2]+FT~\cite{Nason:2009ai,powheg-sm,powheg-kappa-lambda}              & PDFLHC          & \PYTHIA[8.2] & A14 \\
VBF $HH$         & \MGNLO                                                                           & \NNPDF[3.0nlo]  & \PYTHIA[8.2] & A14 \\
\bottomrule
\end{tabular}
}
\end{table}

The backgrounds can be divided into three categories. The largest background category consists of events
with two photon candidates featuring a smoothly falling, non-resonant diphoton mass spectrum. This population
arises from processes with two prompt photons or from jet processes where one or both the photon candidates are
misidentified jets via instrumental effects. The processes $Z({\to}q\bar{q}) \gamma \gamma$ and
\ttyy are also included in this category. All these backgrounds together are referred to as \lq non-resonant diphoton
background\rq, the category is denoted \yy+jets for short but does include $\gamma$+jets and dijet
processes. Due to the significant contribution from instrumental background in the \yy+jets category, its normalisation
is constrained with data using dedicated control regions detailed in Section~\ref{subsec:bkg:yyjets}.

The processes \yy+jets and $Z({\to}q\bar{q}) \gamma \gamma$, with two real photons, are simulated with
\SHERPA[2.2.4] and \SHERPA[2.2.11]~\cite{Bothmann:2019yzt} respectively. Matrix elements at
next-to-leading-order (NLO) in QCD for up to one parton and at LO for up to three partons are calculated
with the Comix~\cite{Gleisberg:2008fv} and \OPENLOOPS~\cite{Buccioni:2019sur,Cascioli:2011va,Denner:2016kdg}
libraries. An alternative \yy+jets MC sample generated with \MGNLO~\cite{Alwall:2014hca} including production
of up to two jets at NLO is considered for evaluation of systematic uncertainties. The \ttyy\ process is
generated with \MGNLO.

The second largest background category consists of processes with photons from a single Higgs boson decay,
this includes processes where the Higgs boson may be produced in association with other particles. This includes single
Higgs boson production via gluon--gluon fusion (ggF), vector-boson fusion (VBF), $WH$, $ZH$ ($qq \to ZH$
and $gg \to ZH$), $t\bar{t}H$, $b\bar{b}H$, $tHq$ and $tHW$. The cross sections of the single Higgs boson
processes are set to the most precise available theoretical values~\cite{deFlorian:2016spz}.

The last background category consists of Standard Model Higgs boson pair production ($HH$) processes via ggF and VBF.
The ggF Higgs boson pair production cross section is calculated at next-to-next-to-leading-order (NNLO) accuracy
including finite top-quark mass effects~\cite{DiMicco:2019ngk,Shao:2013bz,deFlorian:2015moa,Baglio:2020wgt}.
The cross section for Higgs boson pair production via VBF is calculated at
next-to-next-to-next-to-leading-order (N$^{3}$LO)~\cite{DiMicco:2019ngk}. The analysis assumes a branching
ratio of 0.227\% for the Higgs boson decay into two photons and a branching ratio of 58.2\% for the
Higgs boson decay into two $b$-quarks~\cite{deFlorian:2016spz,Djouadi:1997yw}.

The samples use \evtgen~\cite{Lange:2001uf} for the modelling of $b$- and $c$-hadron decays. A full simulation
of the ATLAS detector~\cite{SOFT-2010-01} based on \GEANT~\cite{Agostinelli:2002hh} is used to reproduce the
detector response for single Higgs boson processes. The samples for signal, non-resonant photon production and
Standard Model Higgs boson pair production are processed with the fast simulation~\atlfasttwo~\cite{ATLAS:1300517}
which employs \GEANT except for a parameterisation of the calorimeter response.

Varying numbers of minimum-bias interactions produced with \PYTHIA[8.186]~\cite{Sjostrand:2007gs} using the
\NNPDF[2.3LO] PDF set with the A3 tune~\cite{ATL-PHYS-PUB-2016-017} are overlaid on the hard-scattering event
of all samples to simulate the effect of multiple \pp\ interactions (pile-up) in the same or nearby bunch crossings.
The events are reweighted as a function of the number of interactions per bunch crossing to match the distribution in data.


%

%
\section{Object definitions}
\label{sec:object_definitions}


\label{sec:objdef}
Events are required to have at least one reconstructed collision vertex, defined as a vertex associated with
at least two tracks with transverse momentum (\pt) larger than 0.5~\GeV. The primary vertex is selected from
the reconstructed collision vertices using a neural network algorithm~\cite{HIGG-2013-08} based on extrapolated
photon trajectories and the tracks associated with each candidate vertex.

Photons are reconstructed from topologically connected clusters~\cite{EGAM-2018-01} of energy deposits
in the electromagnetic calorimeter in the region $|\eta| < $ 2.37, excluding the transition region between the
barrel and endcap calorimeters 1.37 $< |\eta| < $ 1.52. Photon candidates are classified as converted or
unconverted based on whether or not they can be associated to conversion vertices or tracks consistent with
photon conversions.

The calibration of the photon energy is based on a multivariate regression algorithm trained with MC samples,
where the simulated distributions of the input variables are corrected with data-driven techniques. The calibrated energy is brought to the absolute scale
by applying scale factors derived from $Z \rightarrow e^{+} e^{-}$ events~\cite{EGAM-2018-01}. The photon
direction is reconstructed using the longitudinal development of the shower in the calorimeters constrained to the luminous region of the proton-beam collisions. In the case of converted photons the
information about the position of the conversion vertex and the tracks associated with conversion is also used.

Photon identification is based on the lateral shower profile of the energy deposits in the first and second
electromagnetic calorimeter layers and on the energy leakage fraction into the hadronic calorimeter~\cite{EGAM-2018-01}.
It reduces the misidentification of hadronic jets containing large neutral components, primarily neutral pions,
which decay into a pair of highly collimated photons. Identification criteria are tuned for converted and
unconverted photons separately, and the {\em Tight} criteria defined in Ref.~\cite{EGAM-2018-01} are applied.

To improve rejection of misidentified photons, two isolation variables are defined to quantify the amount
of activity around a photon. Calorimeter-based isolation $E_\mathrm{T}^\mathrm{iso}$ is defined as the sum
of the transverse energy of topological clusters within a cone of size $\Delta R = \sqrt{\Delta \eta^2 + \Delta \phi^2} $ = 0.2 around the photon,
after first correcting for the energy of the photon candidate itself and for an average expected pile-up
contribution. Track-based isolation $p_\mathrm{T}^\mathrm{iso}$ is defined as the scalar sum of the transverse
momenta of all tracks with $p_\mathrm{T} > $ 1~\GeV\ originating from the primary vertex and within a cone
of size $\Delta R = $ 0.2 around the photon. To be considered isolated a photon must have
$E_\mathrm{T}^\mathrm{iso} / E_\mathrm{T} < 0.065 $ and $p_\mathrm{T}^\mathrm{iso} / E_\mathrm{T} < 0.05$. For isolated photons with transverse energies
between 30~\GeV\ and 250~\GeV, the identification efficiency ranges from
84\% to 98\%~\cite{EGAM-2018-01}.

Electrons are reconstructed from energy deposits measured in the electromagnetic calorimeter that are matched
to ID tracks~\cite{EGAM-2018-01}. They are required to be in the region $|\eta| < $ 2.37, excluding the
transition region between the barrel and endcap calorimeters 1.37 $< |\eta| < $ 1.52, and to have \pt $>$ 10~\GeV.
Electrons are required to satisfy a {\em Medium} identification criterion based on the shower shape, track-cluster
matching and TRT information in a likelihood-based algorithm~\cite{EGAM-2018-01}. Muons are reconstructed from
high-quality tracks found in the MS~\cite{PERF-2015-10}. A matching of the MS tracks to ID tracks is required
in the region $|\eta| < 2.5$. Muons are required to have $|\eta| < 2.7$ and \pt $>$ 10~\GeV\ and to satisfy a
{\em Medium} identification criterion~\cite{MUON-2018-03}. Electrons and muons are both matched to the primary
vertex via requirements on the longitudinal and transverse impact parameters on the tracks, $|z_0|$ and $|d_0|$, respectively.
These requirements are $|z_0|$ sin $\theta$ $<$ 0.5 mm, and $|d_0| / \sigma_{d_{0}} < $ 5 (3) for electrons (muons).

Reconstructed jets are based on particle-flow objects built from noise-suppressed positive-energy topological
clusters in the calorimeter and reconstructed tracks~\cite{PERF-2015-09}. The anti-$k_t$
algorithm~\cite{Cacciari:2008gp, Fastjet} with radius parameter $R = 0.4$ is used. The jet energy is calibrated
by applying several simulation-based corrections and techniques correcting for differences between simulation
and data~\cite{JETM-2018-05}. Jets are required to have rapidity $|y| <$~4.4 and \pt $>$ 25~\GeV. To suppress
jets produced in pile-up interactions, each jet within the tracking acceptance of $|\eta| < $ 2.4 and with
\pt $<$ 60~\GeV\ is required to satisfy the {\em Tight} jet-vertex tagger~\cite{PERF-2014-03} criteria used to
identify jets from the selected primary vertex.

The flavour of jets is determined using a deep-learning neural network, DL1r~\cite{FTAG-2019-07}. The DL1r
$b$-tagging is based on distinctive features of $b$-hadron decays in terms of the impact parameters of
the tracks and the displaced vertices reconstructed in the ID. The inputs of the DL1r network also include
discriminating variables constructed by a recurrent neural network (RNNIP)~\cite{ATL-PHYS-PUB-2017-003},
which exploits the spatial and kinematic correlations between tracks originating from the same $b$-hadron.
For each jet, DL1r gives three different probabilities $p_\mathrm{b}$-, $p_\mathrm{c}$- and
$p_\mathrm{\mathrm{light}}$ for the jet to originate from a $b$, $c$ or light quark respectively. The three
probabilities are combined to define the final discriminant.
The DL1r algorithm is optimised to maximise performance on particle-flow jets and extends the algorithm
performance to very high jet \pt. Only central jets with $|\eta| < $ 2.5 are considered for flavour tagging.
Working points are defined by single requirement values on the DL1r discriminant output distribution, and
can be chosen to provide a specific $b$-jet efficiency for an inclusive $t\bar{t}$ MC sample. The analysis
makes use of the DL1r working point with a 77\% efficiency to select jets containing $b$-hadrons in
simulated $t \bar{t}$ events. For this working point the misidentification rate is 1/130 for light-flavour
jets and 1/4.9 for charm jets.
Scale factors are applied to correct for differences in $b$-tagging efficiency between data and simulation. The
scale factors are measured as a function of the jet \pt using a likelihood-based method in a sample enriched in
$t\bar{t}$ events~\cite{FTAG-2018-01}.
Scale factors are applied to correct for differences in $b$-tagging efficiency between data and simulation. The
scale factors for jets originating from a $b$ quark are measured as a function of the jet \pt using a likelihood-based
method in a sample enriched in $t\bar{t}$ events~\cite{FTAG-2018-01}. The scale factors for jets originating from
a $c$ quark are measured in $t\bar{t}$ events containing a $W\to cs$ decay~\cite{FTAG-2020-08}. For light flavour jets,
the scale factors are derived using $Z$+jets events~\cite{FTAG-2019-02}.

The energy of $b$-tagged jets is corrected for the possible contribution of muons from semileptonic $b$-hadron decays.
Additionally, the undetected energy of neutrinos and out-of-cone effects are corrected for with scale factors derived
as a function of the $b$-jet \pt from a $t\bar{t}$ MC sample. The two corrections together improve the resolution
of the invariant mass of the two jets with the highest $b$-tagging discriminant by about 20\%. This procedure closely follows
that in Ref.~\cite{HIGG-2016-29}.

Overlap removal procedures are applied to avoid using the same detector signals to reconstruct multiple objects. In this
analysis priority is given to photons, by removing jets, electrons and muons within $\Delta R <$~0.4 of a selected photon.
Next, jets within $\Delta R<$~0.2 of electrons are removed. Finally, electrons and muons within $\Delta R <$~0.4 of any remaining
jet are removed.



%

%
\section{Analysis strategy}
\label{sec:event_selection}

A wide range of masses is considered for the scalars \X\ and \S, leading to significantly varying event
kinematics at different hypothetical values of \mX and \mS. When \mX $\gg$ \mS + \mH, the scalar \S\ can become so boosted
that its decay products, two $b$-quarks, become very collimated and are reconstructed within the same $R =$ 0.4 jet.
For smaller values of $\mX - (\mS + \mH)$, two separate $b$-tagged jets are reconstructed. Therefore two mutually exclusive regions are defined with either one or two $b$-tagged jets, referred to as the 1 $b$-tagged and 2 $b$-tagged
regions, and respectively dedicated to the boosted and non-boosted scenarios. A preselection, common to both regions, is introduced below. A final discriminating variable is then defined for each signal hypothesis using parameterised neural networks. The final experimental constraint on the signal
is obtained from a signal-plus-background fit to the distribution of the discriminating variable in data. In the special case where
\mS=125~\GeV, this analysis strategy can be compared with an alternative strategy similar to that applied in the ATLAS search for
$HH \to \ensuremath{b\bar b \gamma \gamma}$~\cite{PhysRevD.106.052001}, and it was found that for any given signal point
the difference between estimated upper limits on $\sigma(\X\to\S\hsm\to b\bar{b}\gamma\gamma)$ is small.

\subsection{Event preselection}

%

Events are selected using diphoton triggers described in Section~\ref{sec:samples}. Beyond the trigger requirements, events are selected if:
\begin{itemize}
\item{At least two photons satisfy the requirements in Section~\ref{sec:objdef}.}
\item{The invariant mass of the two leading photons satisfies 105~$<$ \myy $<$ 160~\GeV.}
\item{The leading photon has \pt$>$~0.35\myy and the subleading photon has \pt$>$~0.25\myy.}
\item{No electrons or muons, as defined in Section~\ref{sec:objdef}, are present.}
\item{The number of central ($|\eta| < $ 2.5) jets is at least two and no more than five. This reduces the \ttH background where top quarks decay hadronically.}
\item{There is exactly one or two $b$-tagged jet at the 77\% working point. Events with more than two $b$-tagged jets are removed to ensure orthogonality with the $b\bar{b}b\bar{b}$ final state from the same signal.}
\end{itemize}


%

\subsection{Signal region definitions}

%
The number of $b$-tagged jets is used to categorise events in two regions, requiring 1 or 2 $b$-tagged jets.
The signal events contain the characteristic $\Htoyy$ decay with the $\myy$ distribution peaking around the Higgs boson
mass at $\sim$125~\GeV. Therefore, two mutually exclusive signal regions (SR) dedicated respectively to the boosted and non-boosted scenarios are defined requiring 120 $< \myy <$ 130~\GeV. The shape of the final signal-to-background discriminant, defined in Section \ref{sec:discriminant} for each signal hypothesis, and obtained in the signal region, is used for the final statistical test of the signal-plus-background and background only hypothesis.

Events with \myy outside the [120, 130]~\GeV interval are instead used to construct the sideband control regions for background estimation as described in
Section~\ref{subsec:background_estimation}. The 2 $b$-tagged and 1 $b$-tagged sideband regions (SB) are found to contain about
85\% of \yy+jets with two real photons and are expected to contain less than 0.1\% of Higgs boson processes.

The fraction of signal events with two resolved $b$-tagged jets is below 50\% if \mS/\mX~$<$~0.09, as derived from simulations.
The 1~$b$-tagged signal region is thus used to analyse signal points for which $\mS/\mX~\lesssim$~0.09 and, conversely, the 2~$b$-tagged
selection is used to analyse signal points for which \mS/\mX~$\gtrsim$~0.09.


%

\subsection{Final signal-to-background discriminant}

\label{sec:discriminant}
Multivariate discriminants are used to separate signal
from background events in the signal regions. Two distinct parameterised neural networks (PNNs)~\cite{Baldi_2016} are trained with events in the 2 $b$-tagged
and the 1 $b$-tagged signal and sideband regions. PNNs take as input a vector of event characteristics $\bar{x}$ and a vector of phase space parameters $\bar{\theta}$
and yield a response function that is parameterised in $\bar{\theta}$.
The parameterisation provides a unique discriminant for each signal hypothesis, separating the targeted signal events from background events. Therefore, for each value of $\bar{\theta} = (\mS, \mX)$, the PNN($\bar{\theta}$) is effectively a different observable.
The PNNs provide sensitivity over the
considered mass range and allow interpolation to values of $\bar{\theta}$ not explicitly included in the training.
In the 2 $b$-tagged signal region the PNN is parameterised in the plane of the two particle masses $\bar{\theta} = (\mS, \mX)$,
and as a function of $\bar{\theta} = (\mX)$ in the 1 $b$-tagged region.

The decay chain $\X\to \Sbb \Hyy$ and the masses \mS and \mX are encoded in the invariant mass of the final
state particles, thus the most effective features to train the PNNs are the invariant masses of the final
state photons and $b$-tagged jets.

For the 2 $b$-tagged signal region the input features are $\bar{x} = (\mbb, \mbbyystar)$ where
\( \mbbyystar = \mbbyy - (\myy - 125 \: \mathrm{GeV}) \). The replacement of \myy by the Higgs boson
mass of 125~\GeV\ allows to remove correlations between the PNN score and \myy, allowing to create sideband regions
for background normalisation as described in Section~\ref{subsec:bkg:yyjets}. For the 1 $b$-tagged signal region
the input variables are $\bar{x} = (\ptb, \mbyystar)$ where \ptb is the \pt of the $b$-tagged jet, and \mbyystar
is derived from the invariant mass \mbyy of the only available $b$-tagged jet and the two photons as
\( \mbyystar = \mbyy - (\myy - 125 \: \mathrm{GeV})  \). Additional variables were considered in the training
but did not bring significant improvements and were therefore disregarded. The usage of the variables
$\mbb$ and $\mbbyystar$ alone allows for the signal interpolation applied in the 2 $b$-tagged signal
region and described in Section~\ref{subsec:interpol}.

The PNNs are trained using Keras~\cite{chollet2015keras} with the Tensorflow~\cite{tensorflow2015-whitepaper} backend.
The training is performed using 69 simulated signal samples chosen from the entire investigated mass grid, as well as
the largest background processes:  non-resonant diphoton+jets, $\ttH$, $ZH$ and
ggF~$H$. In the 1 $b$-tagged signal region the VBF $H$, and Higgs boson pair production processes are also considered for training. As the
vector of parameters $\bar{\theta}$ is not meaningful for the background samples, each background event has $\bar{\theta}$ values
assigned at random from the distribution of values in the signal samples during training.

After the signal region selections, most events arise from the $\yy$ + jets background category,
leading to very unbalanced training classes, which makes it difficult for the PNNs to differentiate
between signal and background. The imbalance is reduced by giving a unit weight to all MC events used
in the training. The effect of using a unit event weight on the shape of the input features $\bar{x}$ is found to be negligible.

The PNN internal architectures are optimised using KerasTuner~\cite{chollet2015keras}, which chooses the hyper-parameters maximizing the Area-Under-Curve
calculated on an evaluation set and using Bayesian Optimisation~\cite{garnett_bayesoptbook_2023}. The class weight defined as
\( w_\mathrm{c} = 0.5 n_{\mathrm{tot}} / n_{\mathrm{c}} \) is used,
where $c$ is either signal or background, $n_{\mathrm{c}}$ is the number of events in the given class and $n_{\mathrm{tot}}$
is the total number of events. Furthermore, given the number $n_s$ ($n_b$) of signal (background) events,
an initial bias of $\mathrm{log}(n_s / n_b)$ is applied to the last layer of the PNN.

The PNNs use binary cross entropy as the loss function and stochastic gradient-based optimisation using the Adam algorithm~\cite{AdamArxiv2014}.
All hidden layers have a standard dense training layer using a rectified linear unit activation function and each has a dropout layer with a dropout rate between 2\% and 20\%.
The output layers have a single node and use a sigmoid activation function. The PNN used in the 2 $b$-tagged signal region is trained with
signal samples with \mX~$\geq$~170~\GeV\ and \mS~$\geq$~30~\GeV. For the 1 $b$-tagged signal region, only signal points where \S\ has enough
boost are used, this includes eleven points with 15~$\leq$\mS$\leq$70~\GeV. The PNN of the 2 $b$-tagged signal region has four hidden
layers with 85, 49, 45 and 81 nodes. The PNN of the 1 $b$-tagged signal region has three hidden layers with 101, 29 and 101 nodes.
After training, the PNN output shape is compared between data and MC in dedicated sideband regions to validate the modelling of the PNN distribution. The results of these comparisons are covered in Section~\ref{sec:crval}.

\subsection{Signal interpolation}
\label{subsec:interpol}
To set continuous limits in the (\mX, \mS) plane, it is desirable to set limits on intermediate
signal models where no sample is simulated. In order to constrain an intermediate signal point
defined by ($\mX^{\mathrm{int}}$, $\mS^{\mathrm{int}}$), the shape of the PNN output is interpolated
from a nearby reference signal sample simulated with low statistical uncertainty and referred to as
($\mX^{\mathrm{ref}}$, $\mS^{\mathrm{ref}}$). The distributions of the input features for ($\mX^{\mathrm{int}}$,
$\mS^{\mathrm{int}}$) are derived from the reference sample in two steps: first a rescaling step that takes
into account the different masses at the reference and intermediate points, and second a reweighting
step that takes different mass resolutions at different mass points into account. The interpolated input features are then used as input to the PNN to obtain the PNN($\theta)$ distribution for the interpolated signal.

For each selected event in the reference sample, the four-vectors of \hsm, \S\ and \X\ are measured using the
selected $b$-jets and photons. The four-vectors of \hsm\ and \S\ are recomputed in the rest frame of \X\
using a Lorentz transformation defined by the four-vector of \X. In absence of any experimental effects, the
four-vectors of \hsm\ and \S\ in the \X\ rest frame would be set to ideal values defined by the kinematics of
the $\X \to \S\hsm$ two-body decay and the specific values of $\mX$ and $\mS$. In practice, the rest frame
quantities are distributions around their ideal values, while the shape of the distributions are
determined by experimental effects. The kinematics at the intermediate point
($\mX^{\mathrm{int}}$, $\mS^{\mathrm{int}}$) are emulated from
($\mX^{\mathrm{ref}}$, $\mS^{\mathrm{ref}}$) by rescaling the rest frame four-vectors of \hsm\ and \S\ so
that they are distributed around their new ideal theoretical values at
($\mX^{\mathrm{int}}$, $\mS^{\mathrm{int}}$), and events of the reference sample are weighted to reproduce
the expected experimental resolution at the intermediate mass.

The resolution effects are much larger for jets than for photons; therefore, the resolution weighting only considers
the $\mbb$ resolution. The $\mbb$ resolution is measured for the simulated points and modelled with a Bukin
probability~\cite{Bukin}. Each parameter of the Bukin probability becomes a 2D map in the (\mX, \mS) plane. The
values of the Bukin parameters can then be interpolated to any mass point ($\mX^{\mathrm{int}}$, $\mS^{\mathrm{int}}$)
using Delaunay triangulation~\cite{DelaunayInterP}. The final distribution of PNN($\mX^{\mathrm{int}}$, $\mS^{\mathrm{int}}$)
is obtained by feeding the values of \mbb and \mbbyystar to the PNN, after four-vector rescaling and resolution weighting.
This technique works well at high momenta where the change in resolution effects between nearby simulated points is
small. For this reason the interpolation is only applied in parts of the 2 $b$-tagged signal region, defined
by \mX$>$300~\GeV and \mS$>$70~\GeV.

The quality of the PNN shape interpolation is evaluated by studying its impact on the expected upper limits
on $\sigma(\X\to\S\hsm\to b\bar{b}\gamma\gamma)$. The upper limits obtained with the PNN output shape from
simulated signal events are compared with those obtained with the interpolated PNN output shape.
In the domain where the interpolation is applied the expected limits on $\sigma(\X\to\S\hsm\to b\bar{b}\gamma\gamma)$
change by less than 5\% for \mS$\geq$100~\GeV\ when replacing the actual PNN output shape with the interpolated one.

For lower masses (below \mS=70~\GeV) fast changes in resolution mean that the limits obtained from interpolated PNN shapes
can differ by more than 10\% from those obtained with the actual PNN shape. For this reason the interpolation
procedure is not applied at low masses or in the 1 $b$-tagged region. Instead a much finer grid of simulated
signal samples in the (\mX, \mS) plane is used. The required granularity is studied by performing injection
tests and checking the sensitivity of one PNN at a given $\bar{\theta}$ to neighbouring signal samples generated
at a different (\mX, \mS) point. The grid spacing is chosen small enough so that a signal excess at one simulated grid point would
also appear in the PNN output of at least one nearby simulated signal sample.


%


%

%
\section{Background estimation}
\label{sec:background_estimation}

\label{subsec:background_estimation}
The final interpretation in terms of a search for a possible $\X\to \Sbb \Hyy$ signal requires
the ability to predict for each background process the number of events in the signal region,
and the shape of the PNN output. In the following sections several methods are employed: a
data-driven technique to study the composition of the non-resonant diphoton background category
in terms of events with zero, one or two misidentified photons, a data-driven method to
normalise the non-resonant diphoton background category, and finally simulations to evaluate
contributions from irreducible backgrounds.

\subsection{Non-resonant diphoton background}
\label{subsec:bkg:yyjets}

The largest background category is the non-resonant \yy+jets background, which includes
instrumental components such as $\gamma$+jets and dijets where jets are misidentified
as photons. Other contributions to this category are the \ttyy process which
represents less than 2\% (0.5\%) of the background in the 2 $b$-tagged (1 $b$-tagged) signal region,
and the $Z({\to}q\bar{q}) \gamma \gamma$ process, which represents less than 1\% (0.3\%) in the 2
$b$-tagged (1 $b$-tagged) region. These last two processes are estimated directly from simulation.

The fractions of \yy+jets events with zero, one or two misidentified photons and their associated systematic
uncertainties are determined by the double two-dimensional sideband method, a data-driven technique which relies
on several photon identification criteria, already employed in the search for Higgs boson pair production
in the $b\bar{b}\gamma\gamma$ final state~\cite{HDBS-2018-34}, the method is described in Ref.~\cite{STDM-2017-30}
and references therein. The fractions of misidentified photon backgrounds are derived individually in the 1 and 2
$b$-tagged regions. The fraction of events with two real photons is found to be 87\% in the 1 $b$-tagged region
and 84\% in the 2 $b$-tagged region. The data-driven method is also used later to derive the \mbbyy, \myy,
\mbb and PNN distributions for the different components with true or misidentified photons. These distributions
are compared in the sideband regions with the data and the \SHERPA \yy+jets MC sample. No large difference
is found between the shapes of the components, indicating that it is possible to directly build high-statistics
templates from \SHERPA samples without adding contributions from misidentified photons.

Based on these studies, the zero, one
or two misidentified photon components of the \yy+jets background category are modelled directly
with the \SHERPA \yy+jets MC sample, with appropriate systematic uncertainties described in Section~\ref{Sec:systematics}.
The normalisation of the \SHERPA \yy+jets MC sample is fitted to the data in the sideband and signal regions
according to the statistical model described in Section~\ref{subsec:stat_model}, the fit effectively rescales
the \SHERPA \yy+jets MC sample to the sum of the zero, one and two misidentified photon components of the non-resonant
diphoton background in data. The rescaling of the \SHERPA \yy+jets MC sample therefore reflects both misidentified photon
contributions and higher order processes present in data but not in the simulation.
Several systematic uncertainties can affect the ratio of the \yy+jets background
in the signal region over the corresponding sideband as detailed in Section~\ref{sec:systematics}.
In a background-only fit where only the data in the sidebands are considered, this procedure scales the
\SHERPA \yy+jets MC sample by a factor 1.03~$\pm$~0.01 in the 1 $b$-tagged region and a factor 1.26~$\pm$~0.03
in the 2 $b$-tagged region.

The shape of the PNN output for the non-resonant diphoton category is studied in the sideband regions in both data
and simulation, and in the signal regions with simulation.  It is observed that the shapes of the
PNN output for the zero, one or two misidentified photon components are well reproduced by the \SHERPA \yy+jets
MC sample. In the \SHERPA \yy+jets simulation the input variables to the PNN are observed to be consistent within statistical
uncertainties between the signal regions and two narrow sidebands defined by \myy $\in$ [115,120] $\cup$ [130,135]~\GeV.
Finally the modelling of the PNN output by simulation is validated in the full sideband control regions,
by comparing the predicted PNN distributions from \SHERPA \yy+jets simulations, with the observed PNN distributions
in data. Theoretical and experimental systematic uncertainties in the PNN output shapes for the \yy+jets background
category are considered in Section~\ref{sec:systematics}.


\subsection{Higgs boson processes}
%

This category of background consists of all processes that contain single Higgs boson production via ggF or
VBF, $WH$, $ZH$ ($qq \to ZH$ and $gg \to ZH$), $t\bar{t}H$, $b\bar{b}H$, $tHq$ and $tHW$. Higgs boson pair production
via ggF and VBF is also included in this category. Following the strategy in Ref.~\cite{HDBS-2018-34},
the normalisation and shape of these processes are obtained from simulated MC samples generated at NLO, and
normalised using state-of-the-art theoretical cross sections~\cite{deFlorian:2016spz}. Several systematic uncertainties
are assigned to these processes as discussed in Section~\ref{sec:systematics}.



%

%
\section{Systematic uncertainties}
\label{sec:systematics}

\label{Sec:systematics}
Three categories of systematic uncertainties are considered. Experimental systematic uncertainties account for
possible differences between the performance of the detector in simulation and in data, they are applied to
all quantities derived from simulation and are described in Section~\ref{Sec:exp_syst}. Theoretical
systematic uncertainties arise where theoretical inputs are used, such as cross sections, or where effects on shapes from
higher-order corrections should be considered, this is described in Section~\ref{subsec:theoretical_systematics}.
The \yy+jets background is derived from a combination of MC simulation and a data-driven technique, and the
related sytematic uncertainties are detailed in Section~\ref{Sec:diphoton_syst}.

\subsection{Experimental systematic uncertainties}
\label{Sec:exp_syst}
Experimental systematic uncertainties affect selection efficiencies and PNN shapes for all Higgs boson processes and the
$\X\to \Sbb \Hyy$ signal. The effect of experimental systematic uncertainties is also propagated to the PNN shape of the \yy+jets
background. The experimental systematic uncertainties are categorised into four groups, in order of decreasing impact
on the expected upper limits on $\sigma(\X\to\S\hsm\to b\bar{b}\gamma\gamma)$: flavour tagging, photons, jets and pile-up.
The limits on $\sigma(\X\to\S\hsm\to b\bar{b}\gamma\gamma)$ are compared with and without experimental systematic uncertainties.
For \mX\ above $\sim$400 \GeV\ the effect of experimental systematic uncertainties on the limits is less than
1\% but grows to 2\%--20\% at lower \mX values.

Flavour tagging uncertainties~\cite{FTAG-2018-01,FTAG-2019-02,FTAG-2020-08} are the leading source of experimental uncertainties in this search.
They include the uncertainties in the efficiency to $b$-tag a jet containing a $b$-hadron, and the probability to $b$-tag a
jet containing a $c$-hadron or a light-flavour jet by mistake. The combined effect of all
flavour tagging uncertainties is the largest in the 2 $b$-tagged signal region with a 5\% uncertainty in the signal efficiency
for models with low \mS and decreasing to 2\% at high \mS. In the 1 $b$-tagged signal region that uncertainty remains in the
range 0.5\%--1.5\%. The corresponding uncertainty in the predicted number of single and double Higgs boson processes is about 3\%.
Once incorporated into the fit and taking into account the effect on both the yields and the PNN shapes of signal and backgrounds,
flavour tagging uncertainties impact the upper limits by about 6\%.

The category of photon-related uncertainties includes uncertainties from efficiencies of the photon triggers, photon
identification and isolation, and uncertainties in the photon energy resolution and scale. These are computed
in data using data-driven techniques~\cite{TRIG-2018-05,EGAM-2018-01} and are propagated to the simulation-based
estimates of background yields and signal efficiencies. The combined effect of all photon uncertainties on the predicted
signal efficiency and predicted number of events from single and double Higgs boson processes is about 2.5\% in both the signal regions.
Once incorporated into the fit and taking into account the effect on both the yields and the PNN shapes on signal and backgrounds,
these translate into an uncertainty of 2\%--4.5\% in the signal sensitivity.

The category of jet-related uncertainties include uncertainties in the jet energy scale and jet energy resolution. These
are derived using data-driven techniques~\cite{JETM-2018-05} and compared with their counterpart in simulation. These
uncertainties are propagated to the signal efficiency and background estimates. The combined effect of all jet-related
uncertainties is largest in the 2 $b$-tagged signal region with up to 14\% uncertainty in the signal efficiency for models with
low \mS and decreasing to about 2\% at high \mS. In the 1 $b$-tagged signal region that uncertainty remains smaller than 3\%.
The corresponding uncertainty in the predicted number of single and double Higgs boson processes is 5\% (7\%) in the
1~$b$-tagged (2~$b$-tagged) signal region. However the effect of these systematic uncertainties on the PNN shape remains
small, of the order of 1\% in the most signal-like PNN bin. Once incorporated into the fit and taking into account the effect
on both the yields and the PNN shapes of signal and backgrounds, these translate into an uncertainty of about 1.5\% in
the signal sensitivity for \mS$>$~110~\GeV. The impact grows at lower masses where it can reach up to 15\%.

The category of pile-up related uncertainties comes from the reweighting of the MC simulation to the pile-up
profile in data. This results in an uncertainty of at most 1\% in the predicted number of events from Higgs boson processes.
The effect on the signal efficiency is smaller than 1\% in both the signal regions and for most of the parameter space,
except in the 2~$b$-tagged signal region for low \mX where it can reach 1.8\%. Finally the 0.83\% uncertainty in the
measured ATLAS Run 2 integrated luminosity is propagated to all processes normalised with their theoretical cross sections.

\subsection{Theoretical systematic uncertainties}
\label{subsec:theoretical_systematics}
Theoretical systematic uncertainties affect the backgrounds normalised by their theoretical cross sections, but also the signal efficiency, and
the non-resonant diphoton background. The latter is discussed in Section~7.3.

For processes with significant contributions to the 1~$b$-tagged and 2~$b$-tagged signal regions and where the dominant
heavy-flavour production is already taken into account at LO ($ttH$ and $ZH$), several theoretical systematic
uncertainties are considered.
Scale uncertainties due to missing higher-order corrections in the production rates are estimated by varying the
factorisation and renormalisation scales up and down from their nominal values by a factor two, taking the envelope
of these variations. Parton shower uncertainties are evaluated by comparing against alternative samples where
the parton shower is performed with \HERWIG[7.1.5]~\cite{Bahr:2008pv,Bellm:2015jjp}. The resulting bin-to-bin variations of the PNN
output range from a few percent up to 10\% in some bins, while the resulting impact on the exclusion limits is below 1\%.
Effects of the choice of PDF and $\alpha_S$ are estimated by varying them following the prescription in
Ref.~\cite{Butterworth:2015oua}. These systematic uncertainties can lead to up to 5\% bin-to-bin variations in the
shape of the PNN discriminant output, but have a small impact on the limits.

For smaller Higgs boson backgrounds where the dominant heavy flavour production occurs at LO, the inclusive
cross section uncertainties from Ref.~\cite{deFlorian:2016spz} are used along with scale, parton shower, PDF and
$\alpha_S$ uncertainties derived as described above. The same procedure is followed for the $Z(\to q\bar{q}) \gamma \gamma$
process. The final impact of these uncertainties on the results is small.

In single Higgs boson processes where $b$-quarks are not produced at LO (VBF $H$, $WH$, ggF~$H$),
a single 100\% normalisation uncertainty is used. This is motivated by studies of heavy-flavour production in association
with top-quark pairs~\cite{TOPQ-2017-12,TOPQ-2012-16} and of $W$-boson production in association with $b$-jets~\cite{STDM-2012-11}.
These have a very small impact on the final sensitivity, except when searching for models with low \mS, where the 100\%
uncertainty in ggF~$H$ can lead to a 10\% decrease in sensitivity to the signal.

The uncertainties in the Higgs boson decay branching ratios $BR(\hsm\to \gamma\gamma)$ and $BR(\hsm\to b\bar{b})$ propagate to
the background yields, impacting the $\sigma(\X\to\S\hsm\to b\bar{b}\gamma\gamma)$ limits by $\sim$3\% and $\sim$1.5\% respectively.

To better understand the overall impact of theoretical background systematic uncertainties, the limits on
$\sigma(\X\to\S\hsm\to b\bar{b}\gamma\gamma)$ are computed with and without the theoretical background systematic uncertainties described
above. It is observed that the limits worsen by about 4\% when including theoretical background systematic uncertainties.
It can be noted that this is smaller than the effect of systematic uncertainties in the non-resonant diphoton background
described in the next section and smaller than the effect of theoretical signal systematic uncertainties.

Theoretical systematic uncertainties are also considered for the $\X\to \Sbb \Hyy$ signal, both in terms of the shape of
the PNN output and the predicted event yields. They include scale, parton shower, PDF and $\alpha_S$ uncertainties
calculated as described earlier. The impact of these systematic uncertainties on the signal efficiency and the shape of
the PNN output  are taken into account. These uncertainties lead to 2\%--10\% bin-to-bin variations of the PNN output shape,
it is most pronounced for the parton shower uncertainty. The systematic uncertainties obtained by changing to an alternative
PDF and $\alpha_S$ lead to up to 11\% change in the predicted signal yield, particularly at high \mX~$\sim$1000~\GeV. The
impact of theoretical signal uncertainties on the sensitivity to $\sigma(\X\to\S\hsm\to b\bar{b}\gamma\gamma)$ can reach 10\%.
A systematic uncertainty associated with the interpolation of the PNN score in the region \mX$>$300~\GeV\ and \mS$>$70~\GeV\
is estimated by considering a shape systematic uncertainty in the PNN score, derived by varying the parameters of the Bukin
probabilities within their errors. The impact on the exclusion limits at interpolated (\mX,\mS) points is at most 10\%.

\subsection{Systematic uncertainties in the non-resonant diphoton background}
\label{Sec:diphoton_syst}
The normalisation of the non-resonant diphoton background is derived in a fit which includes the sideband control regions.
The following theoretical uncertainties, introduced earlier, affect the PNN output shape and the overall estimated number
of events in the signal and sideband regions: scale uncertainties, choice of PDF, $\alpha_s$ and parton shower. The
uncertainties associated with the scales, PDFs and $\alpha_s$ are calculated using the same methods as described in
Section~\ref{subsec:theoretical_systematics}. The parton shower uncertainties are evaluated by using alternative samples
generated with \MADGRAPH~\cite{Alwall:2014hca} for the parton shower. An additional MC modelling uncertainty is considered
by using an alternative \yy+jets sample generated with \MGNLO. This sample models diphoton production with up to two jets
at NLO. The corresponding uncertainty is found to be the most significant in the analysis; its impact in the 2~$b$-tagged signal region
degrades the expected upper limits on the signal by up to 20\%. This impact is larger for low values of $\mX$, while for
$\mX$ above 600~$\GeV$~the impact is always below 5\%. For the 1 $b$-tagged signal region the modelling uncertainty has
an impact of a few percent on the upper limits, except for two signal points at (500, 30)~\GeV\ and (230, 15)~\GeV\ where
the effect is around 40\% due to statistical fluctuations in the alternative \MADGRAPH samples.


%

%
\section{Results}
\label{sec:results}


%
%
%
\subsection{Statistical model}
\label{subsec:stat_model}
The results of the analysis are obtained from a maximum-likelihood fit of the binned PNN output
distribution, performed simultaneously over a signal region and its corresponding sideband
region. The PNN binning is constructed starting from the rightmost signal-like bin. The size of
this bin is chosen to maximise the signal-to-background ratio and widened until there is at least
one background event. The same procedure is repeated with the next bin, and requiring that the number
of background events is above a certain threshold, in order to avoid bins with large statistical errors.
When the signal-to-background ratio in the bin drops below that of the full un-binned distribution,
the iterative process is stopped and a single background-like bin is constructed for the remaining
distribution at PNN outputs close to zero. The binning is optimised independently for each signal hypothesis. For the 1 $b$-jet selection a similar method is used.

The likelihood function is defined as:

\begin{equation}
\mathcal{L} =  \mathrm{Pois}\left( n_{\mathrm{SB}}\left| \mu_{\mathrm{\yy}} N_{\mathrm{SB}}^{\mathrm{\yy}}(\pmb{\theta}) + \sum_{p} N_{\mathrm{SB}}^{p}(\pmb{\theta}) \right. \right) \cdot{\displaystyle\prod_{i} \mathrm{Pois}\left({n_{\mathrm{SR},i} \left| \mu_{\mathrm{\yy}} N_{\mathrm{SR}}^{\mathrm{\yy}}(\pmb{\theta})f_{i}^{\mathrm{\yy}}(\pmb{\theta}) + \sum_{p} N_{\mathrm{SR}}^{p}(\pmb{\theta}) f_{i}^{p}(\pmb{\theta}) \right. } \right) } \cdot G( \pmb{\theta})
\end{equation}

where the index $p$ runs over physics processes other than \yy+jets, the index $i$ runs over the
bins of the PNN output, $n_{\mathrm{SR},i}$ and $n_{\mathrm{SB}}$ are the observed number of events
in the signal region PNN bin $i$ and in the corresponding sideband region,
$N_{\mathrm{SR}}^{p}(\pmb{\theta})$ is the expected number of events from process $p$ in
the signal region, and $N_{\mathrm{SB}}(\pmb{\theta})$ is the total expected number of
events in the corresponding sideband region. The superscript \yy\ is used for parameters
that specifically apply to the \yy+jets background. The factor $\mu_{\mathrm{\yy}}$ is the free
parameter that fits the \yy+jets normalisation to the data. The function $f^{p}_{i}$ gives the shape
or probability density function (pdf) of the PNN output discriminant for each background or
signal process; therefore $N_{\mathrm{SR}}^{p} f_{i}^{p}$ is the expected number of events of
process $p$ in the PNN bin $i$. Finally $\pmb{\theta}$
is a vector of nuisance parameters, and $G(\pmb{\theta})$ are constrained pdfs for the nuisance parameters.
Correlation of the nuisance parameters across different signal and background components, as well as categories,
is taken into account.

The nominal yields of the single and double Higgs boson background processes are initially set to values from simulation.
The likelihood function includes all the nuisance parameters that describe the systematic uncertainties.
The signal cross section is a free parameter in the fit. The measurement of the parameter of interest is
carried out using a statistical test based on the profile likelihood ratio~\cite{Cowan:2010js}.
In the absence of signal, upper limits on $\sigma(\X\to\S\hsm\to b\bar{b}\gamma\gamma)$ at the 95\% CL are set.
The limits are calculated using the asymptotic formula with a profile-likelihood-ratio-based test
statistic~\cite{Cowan:2010js}, and are based on the $\textrm{CL}_{\textrm{S}}$ method~\cite{Read:2002hq}.
The binning in PNN score was chosen to ensure that the asymptotic approximation would be valid.

\subsection{Control region validation}
\label{sec:crval}
\begin{figure}[tbp]
\centering
\subfloat[]{\includegraphics[width=0.45\textwidth]{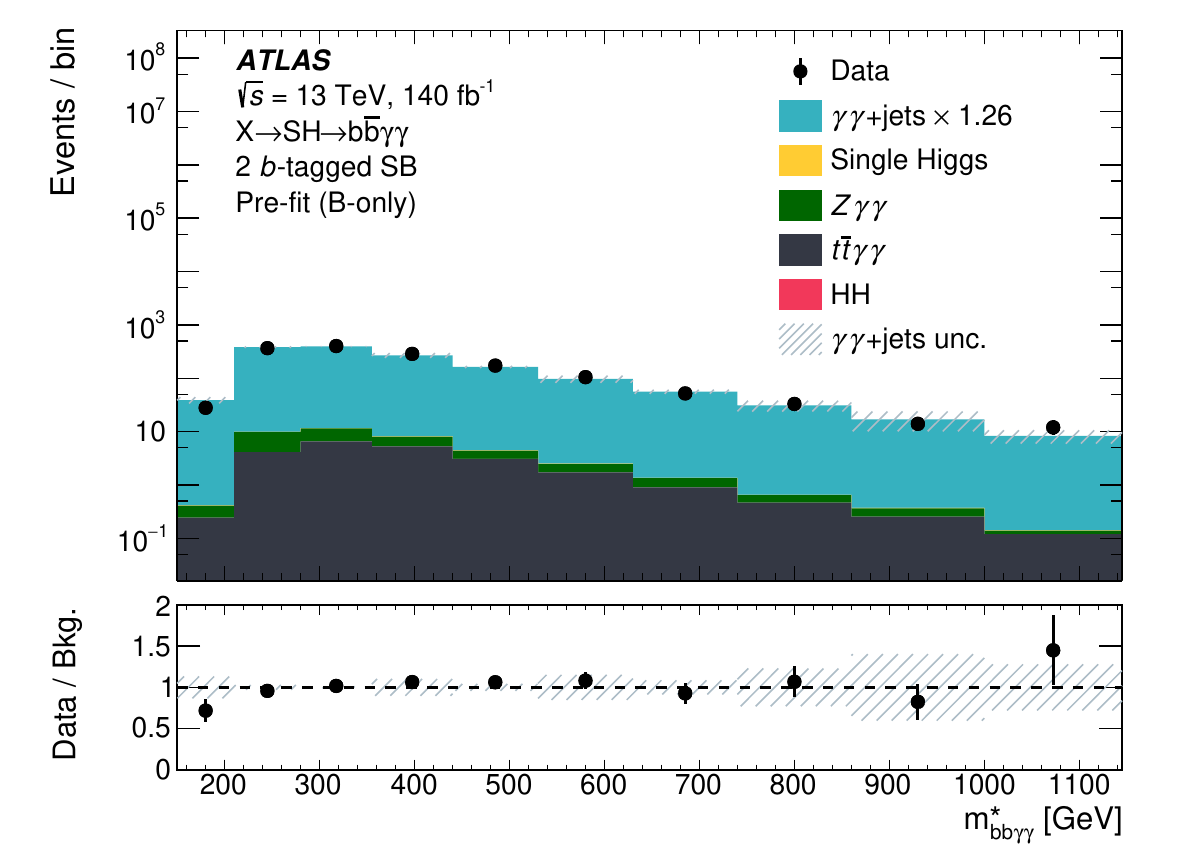}}
\subfloat[]{\includegraphics[width=0.45\textwidth]{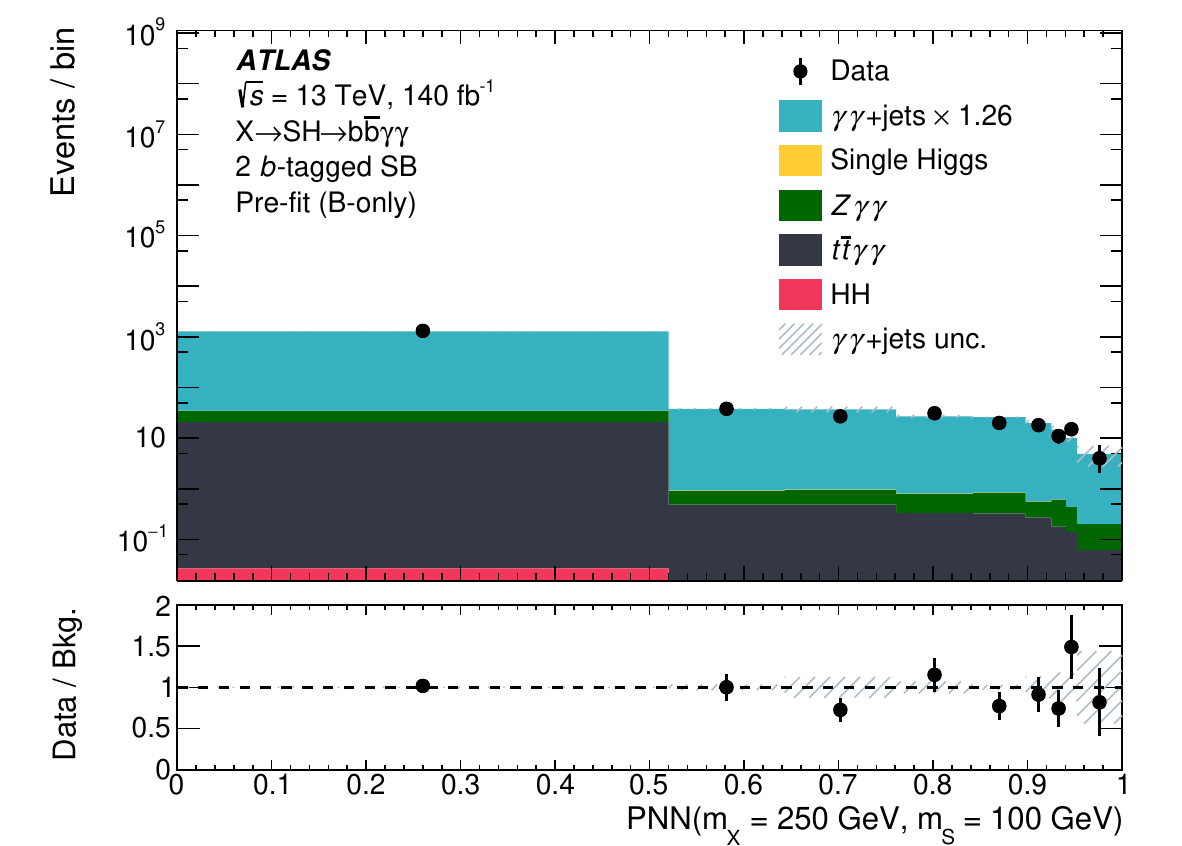}}

\subfloat[]{\includegraphics[width=0.45\textwidth]{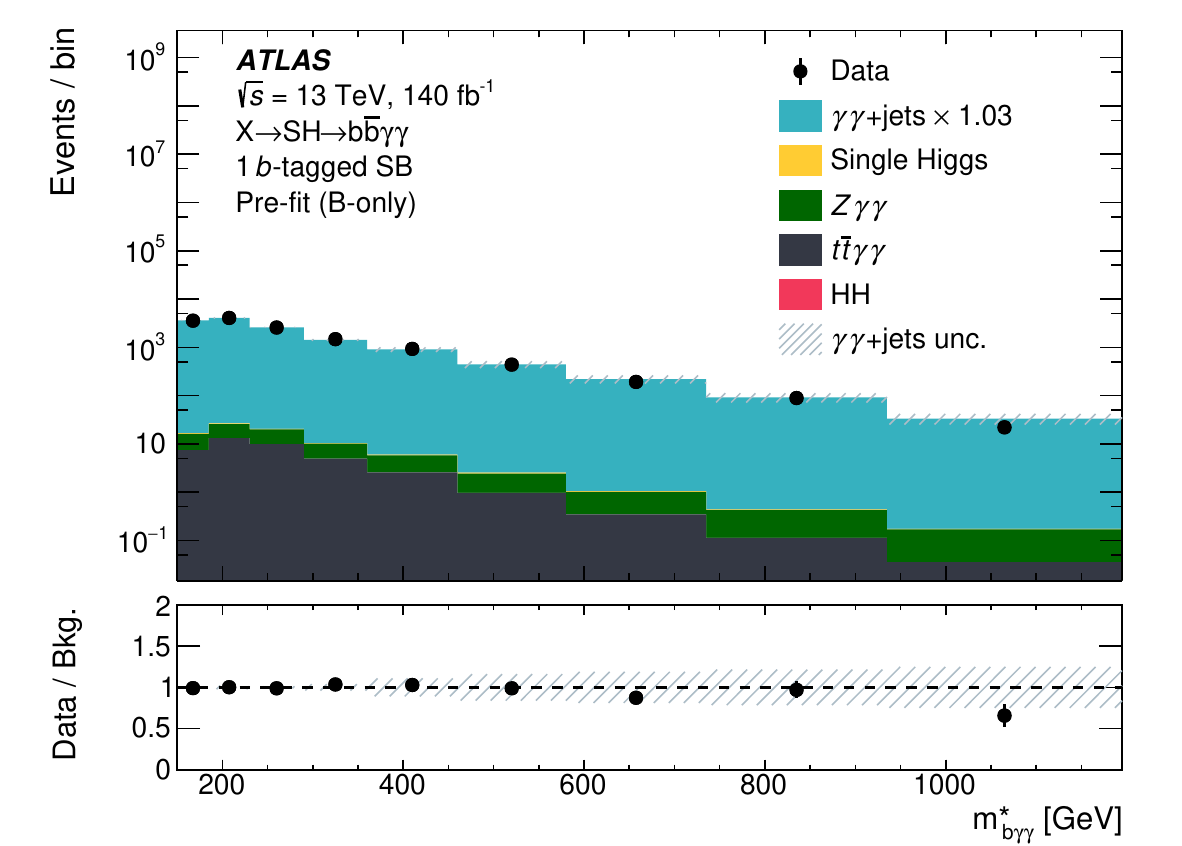}}
\subfloat[]{\includegraphics[width=0.45\textwidth]{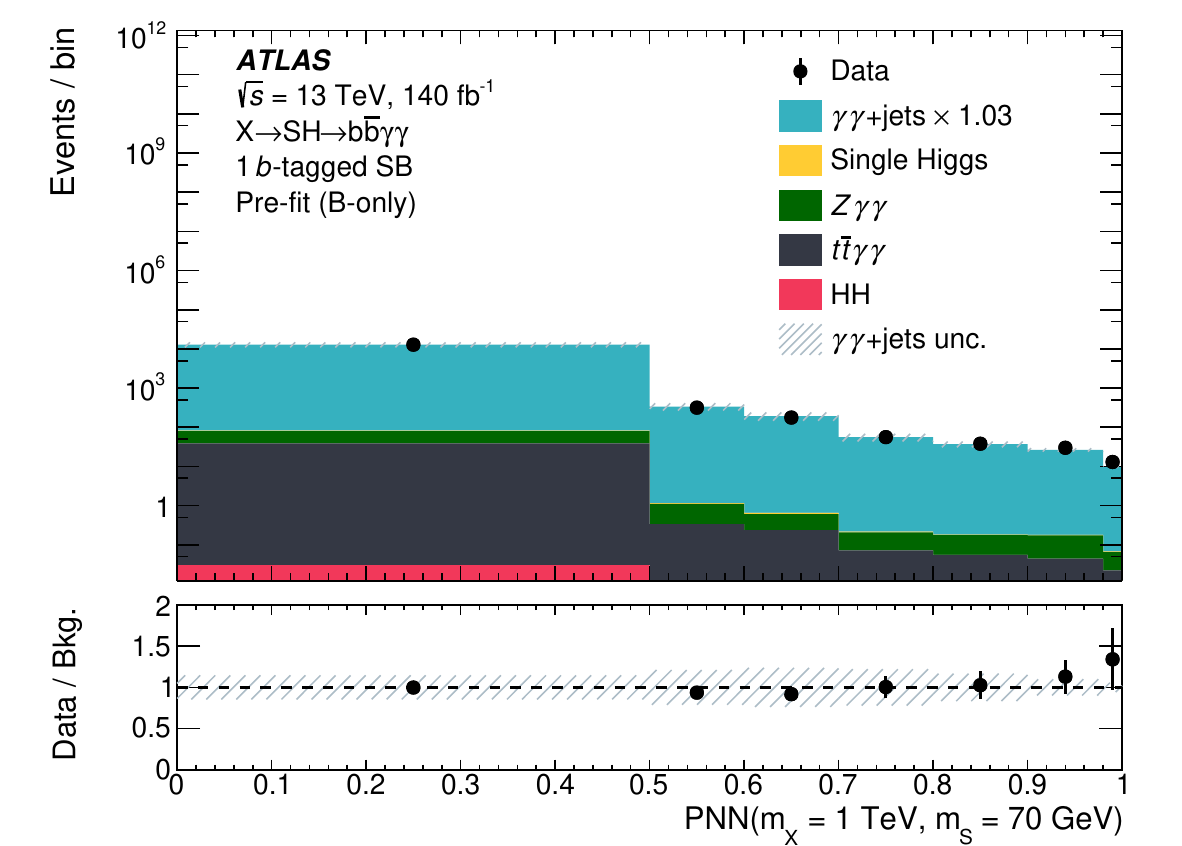}}

\caption{Distributions of (a) \mbbyystar, (c) \mbyystar and (b,d) PNN for two choices of $\bar{\theta} = (\mS, \mX)$ in data and in the predicted model, in the sidebands of the 2 $b$-tagged region (top) and 1 $b$-tagged region (bottom). The \yy~+ jets background is rescaled to its post-fit normalisation in a background-only fit. The variables \mbbyystar and \mbyystar are defined as $\mbbyystar = \mbbyy - (\myy - 125 \: \mathrm{GeV})$ and $\mbyystar = \mbyy - (\myy - 125 \: \mathrm{GeV})$. The \yy~+ jets category represents the sum of \yy~+ jets, $\gamma$ + jets and dijet backgrounds. The error band corresponds to the dominant uncertainty, which arises from the non-resonant $\gamma \gamma$ + jets background.}
\label{fig:validation}
\end{figure}

The ability of the background model to reproduce the data is studied in the sideband regions where a
background-only fit is performed to the data. Distributions are then compared between the post-fit model
and the data as shown in Figure~\ref{fig:validation}. A good agreement between the data and the model shows
the ability of the model to reproduce the PNN discriminant output in both the 2 $b$-tagged and 1 $b$-tagged regions.
For each value of $\bar{\theta} = (\mS, \mX)$ the PNN($\bar{\theta}$) is effectively a different observable,
for this reason the post-fit data-to-prediction comparison in the sideband is performed for all analysed
signal mass points, and good agreement is observed.

\subsection{Result and interpretation}
\begin{figure}[tbp]
\centering
\subfloat[]{\includegraphics[width=0.45\textwidth]{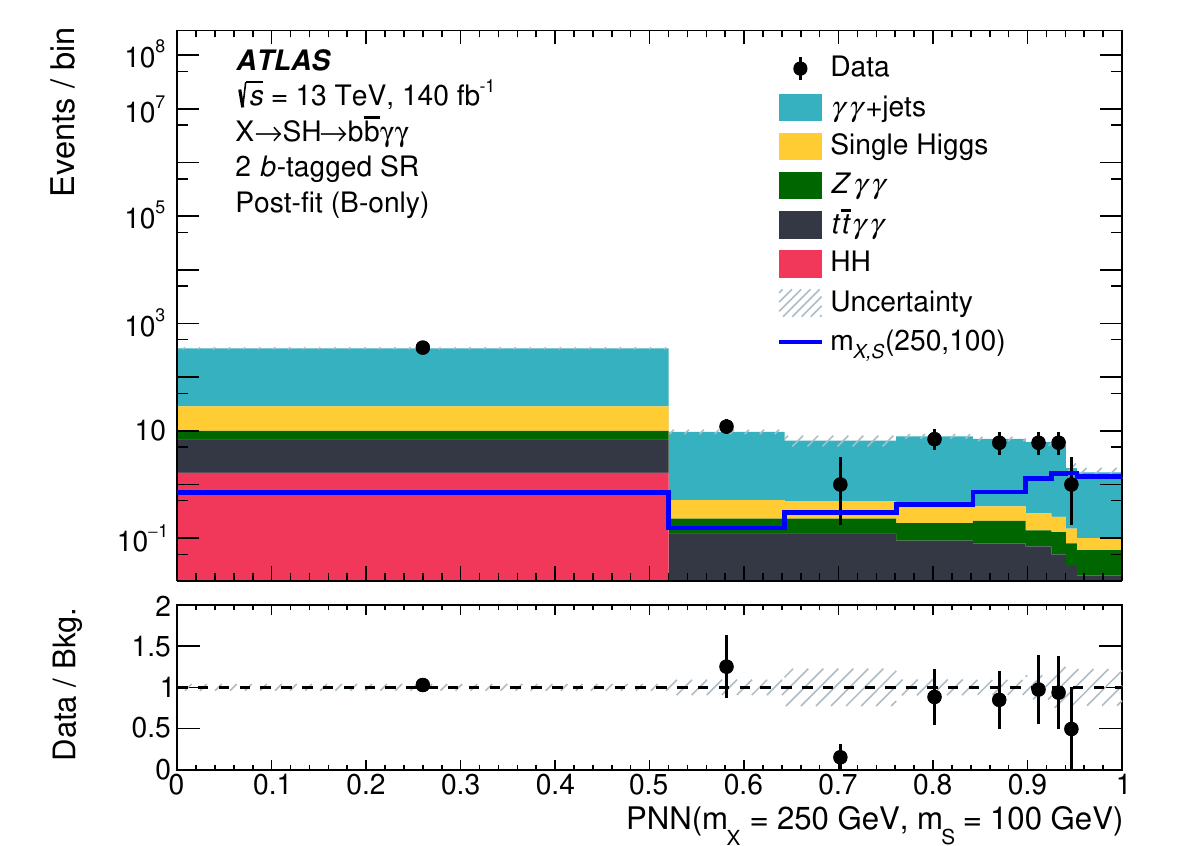}}
\subfloat[]{\includegraphics[width=0.45\textwidth]{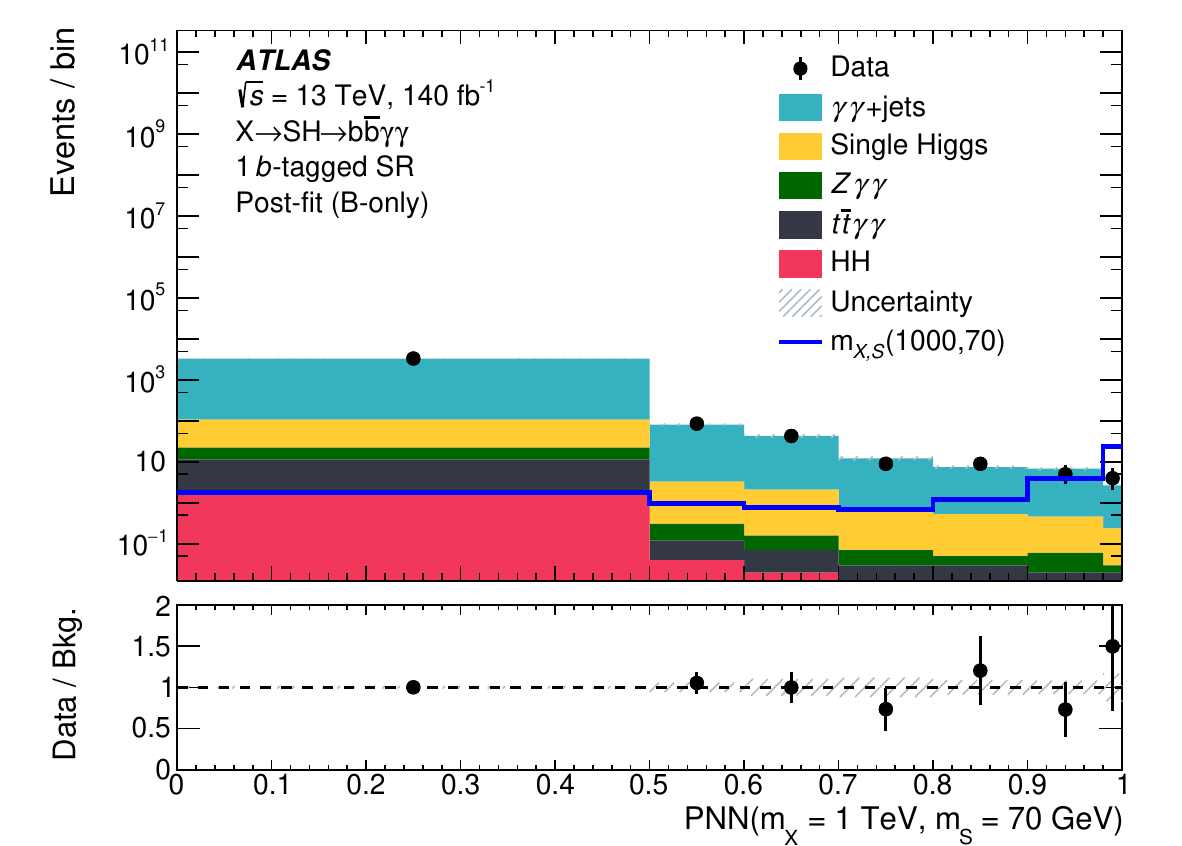}}
\caption{Post-fit distributions of the PNN discriminant output in the (a) 2 $b$-tagged signal region for \mX~=~250~\GeV\ and \mS~=~100~\GeV\ and (b) 1 $b$-tagged signal region for \mX~=~1000~\GeV\ and \mS~=~70~\GeV, after a background-only fit to data. The signals corresponding to the two PNN parameterisations, normalised to a 1~fb cross section, are illustrated for comparison. The \yy + jets category represents the sum of \yy + jets, $\gamma$ + jets and dijet backgrounds. The error band corresponds to the total systematic uncertainty after fit.}
\label{fig:postfit_sr}
\end{figure}

The results of the background-only fit in the signal regions and sidebands to the data is shown
in Table~\ref{tab:bkgfit_to_asimov_sr_sb}. The signal region and sideband yields are independent of the parameterised neural network, while the column "Signal-like bin" illustrates the signal and background yields in the most signal-like bin of the PNN for two choices of $\bar{\theta}$, namely (250, 100)~\GeV~and (1000, 70)~\GeV.
Figure~\ref{fig:postfit_sr} illustrates the post-fit
distribution of the PNN in the 2 $b$-tagged and 1 $b$-tagged regions at two different $\bar{\theta}$.
PNN($\bar{\theta}$) is sensitive not only to a signal at the same masses defined by $\bar{\theta}$, but also
to signals with nearby masses $\bar{\theta'}$. To check for discovery of a wide range of signal
masses, the background-only hypothesis is tested against the data for a highly granular grid of PNN parameters
$\bar{\theta}$. The step size between two consecutive parameters $\bar{\theta}$ is selected by studying the
sensitivity of PNN($\bar{\theta}$) to a signal with different masses $\bar{\theta'}$. If there was an excess
in data due to a signal with masses $\bar{\theta'}$ observed with the discriminant PNN($\bar{\theta}$), the significance of this excess would degrade with increasing distance between $\bar{\theta'}$ and $\bar{\theta}$. A step size in the $\bar{\theta}$ grid is chosen such that the degradation of the significance does not exceed
10\% between two consecutive grid points. In practice the step size goes from 5~\GeV\ in the densest regions to
25~\GeV\ for \mX$\geq$~250~\GeV, and to 50~\GeV~for \mX$\geq$~600~\GeV\ and \mS$\geq$~200~\GeV.

For most mass points good agreement is observed between data and the SM background-only expectation; however
some deviation is observed for a few points. The largest excess of the observation over the SM background-only
hypothesis occurs for $(\mX, \mS)$ = (575, 200)~\GeV\ with a local significance of 3.5$\sigma$. The
\lq look-elsewhere effect\rq\ is taken into account using the asymptotic method described in Ref.~\cite{Gross:2010qma}.
An asymptotic formula for the Euler characteristic as a function of the maximum of the test statistic across each
signal point is derived using toy MC experiments. The resulting global significance is calculated to be 2.0$\sigma$.
At this mass point the signal PNN output shape was initially derived from interpolation. The analysis was repeated
using a simulated sample; the observed significance remained however unchanged, confirming the validity of the
interpolation method.

A parameter point of particular interest is (\mX, \mS) = (650, 90)~\GeV\ where the CMS Collaboration reports
a deviation between observation and background-only expectation, corresponding to a local (global) significance
of 3.8 (2.8) standard deviations~\cite{cms_shbbyy}. Injecting a MC signal with a production cross section of
0.35~fb (the best fit reported by the CMS experiment) in the analysis performed in this paper yields a local
excess in observation to SM expectation of 2.7 standard deviations, demonstrating the sensitivity of this analysis
to a signal consistent with the excess observed by CMS. The results of the analysis of ATLAS data for this
specific parameter point instead shows good agreement between observation and SM background expectation
(the p-value of the background-only hypothesis is larger than~0.5). The 95\%~CL upper limit on
$\sigma(\X\to\S\hsm\to b\bar{b}\gamma\gamma)$ for this specific mass point is 0.2~fb.

\begin{figure}[tbp]
\centering
\subfloat[]{\includegraphics[width=0.45\textwidth]{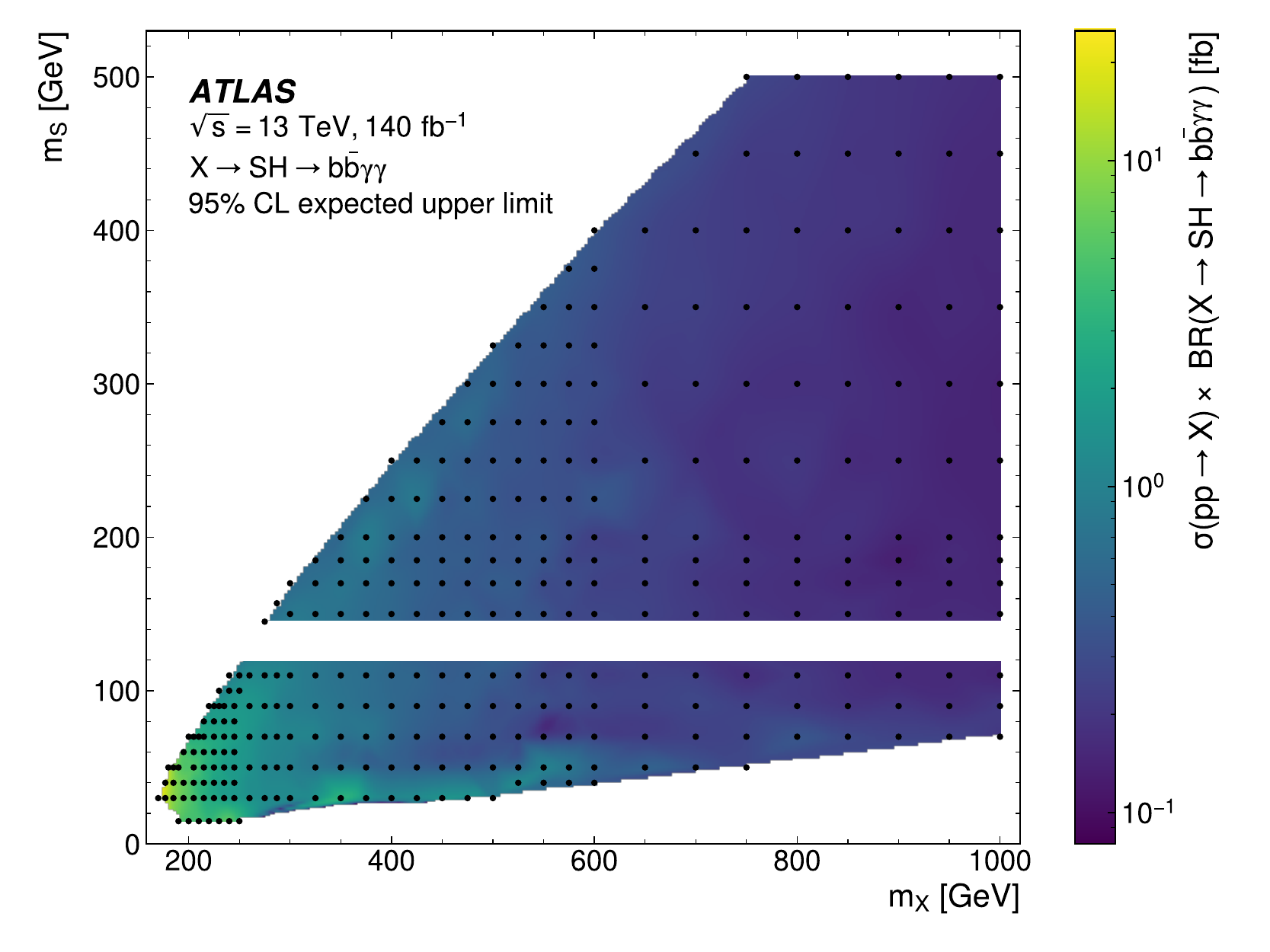}}
\subfloat[]{\includegraphics[width=0.45\textwidth]{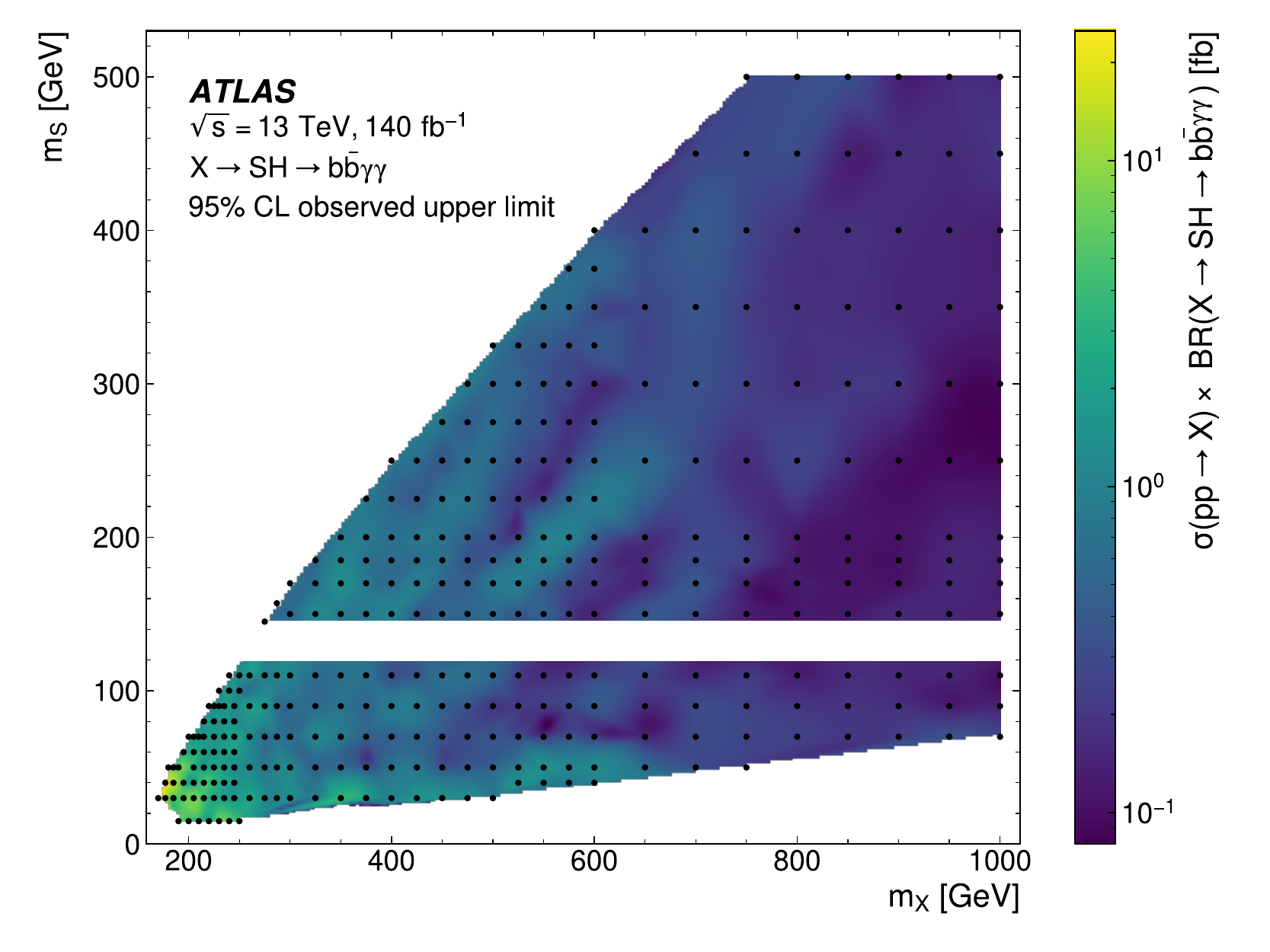}}
\caption{(a) Expected and (b) observed 95\% CL upper limits on the signal cross section times branching fraction for the \mbox{\X $\to$ \S \hsm} signal, in the (\mX, \mS) plane. The points show where the limits were evaluated. The band at \mS~=~125~\GeV~is not shown as those points are equivalent to those already probed in Ref.~\cite{HDBS-2018-34}.}
\label{fig:upper_limit_results}
\end{figure}

The statistical analysis sets 95\% CL upper limits on $\sigma(\X\to\S\hsm\to b\bar{b}\gamma\gamma)$ for all
values of $\bar{\theta} = (\mX, \mS)$ by performing a signal-plus-background fit to the PNN output distribution in data.
For each mass point the signal region which gives the best expected upper limit is selected. The resulting expected and observed
upper limits are presented in Figure~\ref{fig:upper_limit_results}, and range from observed (expected) limits of
39 (25)~fb at \mX = 170~\GeV\ and \mS\ = 30~\GeV,
to 0.09 (0.14)~fb at \mX = 1000~\GeV\ and  \mS\ between 250 and 300~\GeV. The upper limits improve at higher
masses, consistent with the fact that signals with higher \mX become easier to differentiate from SM processes.
In contrast, they worsen in the boosted signal regime at lower \mS, where the 1 $b$-tagged region is employed, and
consistent with the lower signal-to-background ratio shown in Table~\ref{tab:bkgfit_to_asimov_sr_sb} for the 1 $b$-tagged region.
At low \mX the sensitivity suffers from an increasing fraction of $b$-jets falling below the jet \pt reconstruction threshold.

\begin{table*}[tb]
\footnotesize
\caption{Number of events for the different process categories obtained from a background-only fit to data in the signal regions and sidebands. Observed and fitted event yields in the two signal regions and sidebands, independent of phase space parameters $\bar{\theta}$, are given in the columns "Signal region" and "Sideband". The yields in the most signal-like bin of the PNN distribution depend on the selected phase space parameters $\bar{\theta}$, here shown under the column "Signal-like bin" for $\bar{\theta}$ = (\mX, \mS) = (250, 100)~\GeV~and (1000, 70)~\GeV.
The expected number of events from the two corresponding benchmark signals with a 1~fb cross
section is also given. The uncertainties are symmetrised around the central value. The uncertainty in the
total background is calculated taking correlations between the individual contributions into account. For the
single Higgs boson processes, ``Other'' includes the following production modes: VBF, $WH$, $tHq$, and $tHW$.
}
\begin{center}
\begin{tabular}{l | r@{\(\,\pm\,\)}l r@{\(\,\pm\,\)}l r@{\(\,\pm\,\)}l |r@{\(\,\pm\,\)}l r@{\(\,\pm\,\)}l r@{\(\,\pm\,\)}l}
\toprule
\toprule
& \multicolumn{6}{c|}{2 $b$-tagged region} &  \multicolumn{6}{c}{1 $b$-tagged region} \\
\textbf{Background}                   & \multicolumn{2}{c}{\textbf{Sideband}}  & \multicolumn{2}{c}{\textbf{Signal region}} & \multicolumn{2}{c|}{\textbf{Signal-like bin}}          & \multicolumn{2}{c}{\textbf{Sideband}}                  & \multicolumn{2}{c}{\textbf{Signal region}} & \multicolumn{2}{c}{\textbf{Signal-like bin}}   \\
\midrule
Non-res. $\gamma\gamma$                & 1480 & 37   & 372  & 16 & 1.64 & 0.37 & 13450 & 110 & 3392 & 53 & 2.45 & 0.43  \\[4pt]
Single Higgs                           & 0.46 & 0.11 & 19.9 & 5.3 & 0.04 & 0.01 & 2.3 & 1.1 & 92 & 44 & 0.21 & 0.10 \\[4pt]
\phantom{SH}\ggF$+b\bar{b}H$           & 0.14 & 0.11 & 6.5  & 5.2 & 0.01 & 0.01 & 1.5 & 1.1 & 56 & 43 & 0.11 & 0.09   \\[4pt]
\phantom{SH}\ttH                       & 0.21 & 0.01 & 7.91 & 0.77 & 0.01 & 0.01 & 0.31 & 0.01 & 11.4 & 1.1 & 0.03 & 0.01 \\[4pt]
\phantom{SH}$ZH$                       & 0.08 & 0.01 & 3.56 & 0.30 & 0.02 & 0.01 & 0.17 & 0.01 & 7.35 & 0.60 & 0.02 & 0.01 \\[4pt]
\phantom{SH}Other                      & 0.03 & 0.01 & 1.94 & 0.70 & \multicolumn{2}{l|}{$<$ 0.005} & 0.40 & 0.23 & 17 & 10 & 0.05 & 0.03 \\[4pt]
Double Higgs                           & 0.03 & 0.01 & 1.65 & 0.25 & \multicolumn{2}{l|}{$<$ 0.005} & 0.03 & 0.01 & 1.79 & 0.27 & 0.01 & 0.01 \\[4pt]
\midrule
Total 	 			     & 1480 & 37   & 394  & 16   & 1.67 & 0.37 & 13450 & 110 & 3486 & 48 & 2.67 & 0.45 \\[4pt]
\midrule
Signal  (\mX, \mS)                     & \multicolumn{6}{c|}{}               & \multicolumn{6}{c}{}            \\[4pt]
(250, 100) \GeV                        & 0.38 & 0.04 & 8.3 & 1.2 & 1.43 & 0.21 & \multicolumn{6}{c}{} \\[4pt]
(1000, 70) \GeV                        & \multicolumn{6}{c|}{}               & 0.97 & 0.10 & 33.3 & 5.8 & 23.9 & 4.2 \\[4pt]
\midrule
Data                                   & \multicolumn{2}{c}{1479}            & \multicolumn{2}{c}{395} & \multicolumn{2}{c|}{0} & \multicolumn{2}{c}{13450}                      & \multicolumn{2}{c}{3491} & \multicolumn{2}{c}{4} \\
\bottomrule
\bottomrule
\end{tabular}
\label{tab:bkgfit_to_asimov_sr_sb}
\end{center}
\end{table*}

\FloatBarrier


%

%
%
%

%
\section{Conclusion}
\label{sec:conclusion}

A search for a signal from a hypothetical scalar \X\ is performed, considering the case
where it decays into another hypothetical scalar \S\ and a Higgs boson, which subsequently decay into pairs
of $b$-quarks and photons, respectively. Two signal regions targeting resolved or boosted $\Stobb$
decays are analysed using parameterised neural networks, which provide continuous sensitivity in
the probed (\mX, \mS) plane. In the region $\mX>$~300~\GeV\ and $\mS>70$~\GeV\ the validity of the limits for
intermediate mass points is ensured by interpolating the signal shapes to a much finer signal grid, and
the finer grid spacing is guided by the sensitivity range of the PNN to values of (\mX, \mS) where
it was not trained. At lower masses the validity of the limits for intermediate mass points is ensured
by using a very fine grid of simulated signal samples.

No significant excess with respect to the Standard Model background is found. Therefore, 95\% CL upper
limits are set on $\sigma(\X\to\S\hsm\to b\bar{b}\gamma\gamma)$ in the ranges 170~$\leq$~\mX~$\leq$~1000~\GeV\
and 15~$\leq$~\mS~$\leq$~500~\GeV, expanding earlier LHC results to lower masses and providing higher sensitivity.
The largest deviation from the background-only expectation occurs for  $(\mX, \mS) =$\ (575, 200)~\GeV\ with
a local (global) significance of 3.5 (2.0) standard deviations. For the mass point $(\mX, \mS) =$~(650, 90)~\GeV,
where CMS reported an excess with a local (global) significance of 3.8 (2.8) standard deviations, this analysis
shows good agreement with the background-only hypothesis and sets a 95\% CL upper limit on the
signal cross section of 0.2~fb.


%

%
\section*{Acknowledgements}
%

%
%

%
%

We thank CERN for the very successful operation of the LHC and its injectors, as well as the support staff at
CERN and at our institutions worldwide without whom ATLAS could not be operated efficiently.

The crucial computing support from all WLCG partners is acknowledged gratefully, in particular from CERN, the ATLAS Tier-1 facilities at TRIUMF/SFU (Canada), NDGF (Denmark, Norway, Sweden), CC-IN2P3 (France), KIT/GridKA (Germany), INFN-CNAF (Italy), NL-T1 (Netherlands), PIC (Spain), RAL (UK) and BNL (USA), the Tier-2 facilities worldwide and large non-WLCG resource providers. Major contributors of computing resources are listed in Ref.~\cite{ATL-SOFT-PUB-2023-001}.

We gratefully acknowledge the support of ANPCyT, Argentina; YerPhI, Armenia; ARC, Australia; BMWFW and FWF, Austria; ANAS, Azerbaijan; CNPq and FAPESP, Brazil; NSERC, NRC and CFI, Canada; CERN; ANID, Chile; CAS, MOST and NSFC, China; Minciencias, Colombia; MEYS CR, Czech Republic; DNRF and DNSRC, Denmark; IN2P3-CNRS and CEA-DRF/IRFU, France; SRNSFG, Georgia; BMBF, HGF and MPG, Germany; GSRI, Greece; RGC and Hong Kong SAR, China; ISF and Benoziyo Center, Israel; INFN, Italy; MEXT and JSPS, Japan; CNRST, Morocco; NWO, Netherlands; RCN, Norway; MEiN, Poland; FCT, Portugal; MNE/IFA, Romania; MESTD, Serbia; MSSR, Slovakia; ARIS and MVZI, Slovenia; DSI/NRF, South Africa; MICINN, Spain; SRC and Wallenberg Foundation, Sweden; SERI, SNSF and Cantons of Bern and Geneva, Switzerland; NSTC, Taipei; TENMAK, T\"urkiye; STFC/UKRI, United Kingdom; DOE and NSF, United States of America.

Individual groups and members have received support from BCKDF, CANARIE, CRC and DRAC, Canada; CERN-CZ, PRIMUS 21/SCI/017 and UNCE SCI/013, Czech Republic; COST, ERC, ERDF, Horizon 2020, ICSC-NextGenerationEU and Marie Sk{\l}odowska-Curie Actions, European Union; Investissements d'Avenir Labex, Investissements d'Avenir Idex and ANR, France; DFG and AvH Foundation, Germany; Herakleitos, Thales and Aristeia programmes co-financed by EU-ESF and the Greek NSRF, Greece; BSF-NSF and MINERVA, Israel; Norwegian Financial Mechanism 2014-2021, Norway; NCN and NAWA, Poland; La Caixa Banking Foundation, CERCA Programme Generalitat de Catalunya and PROMETEO and GenT Programmes Generalitat Valenciana, Spain; G\"{o}ran Gustafssons Stiftelse, Sweden; The Royal Society and Leverhulme Trust, United Kingdom.

In addition, individual members wish to acknowledge support from CERN: European Organization for Nuclear Research (CERN PJAS); Chile: Agencia Nacional de Investigaci\'on y Desarrollo (FONDECYT 1190886, FONDECYT 1210400, FONDECYT 1230812, FONDECYT 1230987); China: National Natural Science Foundation of China (NSFC - 12175119, NSFC 12275265, NSFC-12075060); Czech Republic: PRIMUS Research Programme (PRIMUS/21/SCI/017); EU: H2020 European Research Council (ERC - 101002463); European Union: European Research Council (ERC - 948254, ERC 101089007), Horizon 2020 Framework Programme (MUCCA - CHIST-ERA-19-XAI-00), European Union, Future Artificial Intelligence Research (FAIR-NextGenerationEU PE00000013), Italian Center for High Performance Computing, Big Data and Quantum Computing (ICSC, NextGenerationEU); France: Agence Nationale de la Recherche (ANR-20-CE31-0013, ANR-21-CE31-0013, ANR-21-CE31-0022, ANR-22-EDIR-0002), Investissements d'Avenir Labex (ANR-11-LABX-0012); Germany: Baden-Württemberg Stiftung (BW Stiftung-Postdoc Eliteprogramme), Deutsche Forschungsgemeinschaft (DFG - 469666862, DFG - CR 312/5-2); Italy: Istituto Nazionale di Fisica Nucleare (ICSC, NextGenerationEU), Ministero dell'Università e della Ricerca (PRIN - 20223N7F8K - PNRR M4.C2.1.1); Japan: Japan Society for the Promotion of Science (JSPS KAKENHI JP21H05085, JSPS KAKENHI JP22H01227, JSPS KAKENHI JP22H04944, JSPS KAKENHI JP22KK0227); Netherlands: Netherlands Organisation for Scientific Research (NWO Veni 2020 - VI.Veni.202.179); Norway: Research Council of Norway (RCN-314472); Poland: Polish National Agency for Academic Exchange (PPN/PPO/2020/1/00002/U/00001), Polish National Science Centre (NCN 2021/42/E/ST2/00350, NCN OPUS nr 2022/47/B/ST2/03059, NCN UMO-2019/34/E/ST2/00393, UMO-2020/37/B/ST2/01043, UMO-2021/40/C/ST2/00187, UMO-2022/47/O/ST2/00148); Slovenia: Slovenian Research Agency (ARIS grant J1-3010); Spain: BBVA Foundation (LEO22-1-603), Generalitat Valenciana (Artemisa, FEDER, IDIFEDER/2018/048), La Caixa Banking Foundation (LCF/BQ/PI20/11760025), Ministry of Science and Innovation (MCIN \& NextGenEU PCI2022-135018-2, MICIN \& FEDER PID2021-125273NB, RYC2019-028510-I, RYC2020-030254-I, RYC2021-031273-I, RYC2022-038164-I), PROMETEO and GenT Programmes Generalitat Valenciana (CIDEGENT/2019/023, CIDEGENT/2019/027); Sweden: Swedish Research Council (VR 2018-00482, VR 2022-03845, VR 2022-04683, VR grant 2021-03651), Knut and Alice Wallenberg Foundation (KAW 2017.0100, KAW 2018.0157, KAW 2018.0458, KAW 2019.0447, KAW 2022.0358); Switzerland: Swiss National Science Foundation (SNSF - PCEFP2\_194658); United Kingdom: Leverhulme Trust (Leverhulme Trust RPG-2020-004); United States of America: U.S. Department of Energy (ECA DE-AC02-76SF00515), Neubauer Family Foundation.

%
%


%
%
%
%
%
%
%

%
%
%
%
%
%

%
%

%
%
%
\printbibliography
\clearpage
\input{atlas_authlist}
%

%
%

%

%
%
%
%

\end{document}

%% file: atlas_authlist.tex
 
\begin{flushleft}
\hypersetup{urlcolor=black}
{\Large The ATLAS Collaboration}

\bigskip

\AtlasOrcid[0000-0002-6665-4934]{G.~Aad}$^\textrm{\scriptsize 103}$,
\AtlasOrcid[0000-0001-7616-1554]{E.~Aakvaag}$^\textrm{\scriptsize 16}$,
\AtlasOrcid[0000-0002-5888-2734]{B.~Abbott}$^\textrm{\scriptsize 121}$,
\AtlasOrcid[0000-0002-1002-1652]{K.~Abeling}$^\textrm{\scriptsize 55}$,
\AtlasOrcid[0000-0001-5763-2760]{N.J.~Abicht}$^\textrm{\scriptsize 49}$,
\AtlasOrcid[0000-0002-8496-9294]{S.H.~Abidi}$^\textrm{\scriptsize 29}$,
\AtlasOrcid[0009-0003-6578-220X]{M.~Aboelela}$^\textrm{\scriptsize 44}$,
\AtlasOrcid[0000-0002-9987-2292]{A.~Aboulhorma}$^\textrm{\scriptsize 35e}$,
\AtlasOrcid[0000-0001-5329-6640]{H.~Abramowicz}$^\textrm{\scriptsize 152}$,
\AtlasOrcid[0000-0002-1599-2896]{H.~Abreu}$^\textrm{\scriptsize 151}$
% ...
\bigskip
\\

% ...
$^{16}$Department for Physics and Technology, University of Bergen, Bergen; Norway.\\
% ...
$^{29}$Physics Department, Brookhaven National Laboratory, Upton NY; United States of America.\\
% ...
$^{35}$$^{(a)}$Facult\'e des Sciences Ain Chock, Universit\'e Hassan II de Casablanca;$^{(b)}$Facult\'{e} des Sciences, Universit\'{e} Ibn-Tofail, K\'{e}nitra;$^{(c)}$Facult\'e des Sciences Semlalia, Universit\'e Cadi Ayyad, LPHEA-Marrakech;$^{(d)}$LPMR, Facult\'e des Sciences, Universit\'e Mohamed Premier, Oujda;$^{(e)}$Facult\'e des sciences, Universit\'e Mohammed V, Rabat;$^{(f)}$Institute of Applied Physics, Mohammed VI Polytechnic University, Ben Guerir; Morocco.\\
% ...
$^{44}$Physics Department, Southern Methodist University, Dallas TX; United States of America.\\
% ...
$^{49}$Fakult\"{a}t Physik , Technische Universit{\"a}t Dortmund, Dortmund; Germany.\\
% ...
$^{55}$II. Physikalisches Institut, Georg-August-Universit\"{a}t G\"ottingen, G\"ottingen; Germany.\\
% ...
$^{103}$CPPM, Aix-Marseille Universit\'e, CNRS/IN2P3, Marseille; France.\\
% ...
$^{121}$Homer L. Dodge Department of Physics and Astronomy, University of Oklahoma, Norman OK; United States of America.\\
% ...
$^{151}$Department of Physics, Technion, Israel Institute of Technology, Haifa; Israel.\\
$^{152}$Raymond and Beverly Sackler School of Physics and Astronomy, Tel Aviv University, Tel Aviv; Israel.\\
% ...
$^{a}$ Also Affiliated with an institute covered by a cooperation agreement with CERN.\\
$^{b}$ Also at An-Najah National University, Nablus; Palestine.\\
$^{c}$ Also at Borough of Manhattan Community College, City University of New York, New York NY; United States of America.\\
$^{d}$ Also at Center for High Energy Physics, Peking University; China.\\
$^{e}$ Also at Center for Interdisciplinary Research and Innovation (CIRI-AUTH), Thessaloniki; Greece.\\
$^{f}$ Also at Centro Studi e Ricerche Enrico Fermi; Italy.\\
% ...

\end{flushleft}
